\pdfoutput=1 

\documentclass[a4paper, 11pt, final]{article}

\usepackage{amssymb} 
\usepackage{amsthm} 
\usepackage{amsmath,empheq}
\usepackage{bbm}

\DeclareMathAlphabet{\mathcal}{OMS}{cmsy}{m}{n}
\DeclareMathAlphabet\mathbfcal{OMS}{cmsy}{b}{n}

\DeclareFontFamily{U}{dutchcal}{\skewchar\font=45 }
\DeclareFontShape{U}{dutchcal}{m}{n}{<-> s*[1.0] dutchcal-r}{}
\DeclareFontShape{U}{dutchcal}{b}{n}{<-> s*[1.0] dutchcal-b}{}
\DeclareMathAlphabet{\mathcald}{U}{dutchcal}{m}{n}
\SetMathAlphabet{\mathcald}{bold}{U}{dutchcal}{b}{n}
\DeclareMathAlphabet\mathcalz{T1}{pzc}{mb}{it}

\usepackage[nice]{nicefrac} 
\usepackage{siunitx} 

\usepackage{newtxtext, newtxmath} 

\usepackage{anyfontsize}
\usepackage[utf8]{inputenc}
\usepackage[T1]{fontenc}
\usepackage{microtype} 

\providecommand{\JEL}[1]{\textit{\textbf{JEL: }} #1}
\providecommand{\keywords}[1]{\textbf{\textit{Keywords--- }} #1}

\usepackage{setspace} 

\usepackage{parskip} 
\usepackage{etoolbox}
\usepackage{titlesec}
\titleformat{\section}{\normalfont\Large\bfseries}{\thesection}{1em}{}
\titleformat{\subsection}{\normalfont\large\bfseries}{\thesubsection.}{1em}{}
\titleformat{\subsubsection}{\normalfont\normalsize\itshape}{\thesubsubsection.}{1em}{}

\usepackage{abstract}

\renewenvironment{abstract}
 {\normalfont
  \begin{center}
  \bfseries \abstractname\vspace{-.5em}\vspace{0pt}
  \end{center}
  \list{}{
    \setlength{\leftmargin}{0cm}%
    \setlength{\rightmargin}{\leftmargin}%
  }%
  \item\relax}
 {\endlist}

\usepackage{authblk} 
\usepackage[bottom]{footmisc} 

\usepackage{multicol} 

\usepackage{enumitem} 

\usepackage{graphicx} 
\usepackage{float} 
\usepackage{subcaption}
\usepackage{afterpage} 

\usepackage{printlen}

\usepackage[labelfont=bf]{caption}
\usepackage{color, colortbl}
\definecolor{LightGray}{rgb}{0.93,0.914,0.914}    
\usepackage{soul} 
\usepackage{longtable,rotating} 
\usepackage{booktabs} 
\usepackage{multirow} 
\usepackage{arydshln} 
\usepackage{bigdelim} 

\usepackage[most]{tcolorbox}            

\PassOptionsToPackage{hyphens}{url} 

\usepackage{fancyhdr}
\fancyheadoffset{0cm}
\makeatletter
\newcommand{\quickwordcount}[1]{
  \immediate\write18{texcount -quiet -incbib -sub=none -utf8 -1 -sum -merge -encoding=utf8 #1.tex > #1-words}%
  \immediate\openin\somefile=#1-words
  \read\somefile to \@@localdummy
  \immediate\closein\somefile
  \setcounter{wordcounter}{\@@localdummy}
  \@@localdummy
}
\makeatother

\usepackage{silence}
\usepackage[style=apa,backend=biber,natbib,hyperref]{biblatex}
\DeclareLanguageMapping{british}{british-apa}
\defbibenvironment{bibliography}{\enumerate}{\endenumerate}{\item}

\usepackage[colorlinks=true,allcolors=gray]{hyperref} 
\let\orgautoref\autoref

\renewcommand{\autoref}[1]
{%
\def\equationautorefname{Eq.}%
\def\sectionautorefname{Sec.}%
\def\subsectionautorefname{Subsec.}%
\def\figureautorefname{Fig.}%
\def\subfigureautorefname{Fig.}%
\orgautoref{#1}%
}

\usepackage[nameinlink,capitalise]{cleveref}

\usepackage{algorithm,algpseudocode}

\makeatletter
\newlength{\trianglerightwidth}
\algnewcommand{\LineCommentCont}[1]{\Statex \hskip\ALG@thistlm%
  \parbox[t]{\dimexpr\linewidth-\ALG@thistlm}
{\leftskip=\algorithmicindent
  \hangindent=\algorithmicindent 
  \hangafter=1%
  \strut\makebox[\algorithmicindent][c]{$\triangleright$}#1\strut}
  } 
\makeatother

\usepackage[left=1.8cm, right=1.8cm, bottom=2.5cm, top=2.5cm]{geometry}

\begin{document}


\renewcommand{\figureautorefname}{Fig.}
\onehalfspacing



\newcommand{\MainTitleText}{
Modelling the term-structure of default risk under IFRS 9 within a multistate regression framework
}

\title{\fontsize{20pt}{0pt}\selectfont\textbf{\MainTitleText
}}


\author[,a,b]{\large Arno Botha \thanks{ ORC iD: 0000-0002-1708-0153; Corresponding author: \url{arno.spasie.botha@gmail.com}}}
\author[,a,b]{\large Tanja Verster \thanks{ ORC iD: 0000-0002-4711-6145; email: \url{tanja.verster@nwu.ac.za}}}
\author[,a]{\large Roland Breedt \thanks{ Email: \url{rolandbreedt00@icloud.com}}}
\affil[a]{\footnotesize \textit{Centre for Business Mathematics and Informatics \& Unit for Data Science and Computing, North-West University, Potchefstroom, South Africa}}
\affil[b]{\footnotesize \textit{National Institute for Theoretical and Computational Sciences (NITheCS), Potchefstroom, South Africa}}
\renewcommand\Authands{, and }

    

\makeatletter
\renewcommand{\@maketitle}{
    \newpage
     \null
     \vskip 1em%
     \begin{center}%
      {\LARGE \@title \par
      	\@author \par
        }
     \end{center}%
     \par
 } 
 \makeatother
 
 \maketitle

{
    \setlength{\parindent}{0cm}
    \rule{1\columnwidth}{0.4pt}
    \begin{abstract}
    The lifetime behaviour of loans is notoriously difficult to model, which can compromise a bank's financial reserves against future losses, if modelled poorly. Therefore, we present a data-driven comparative study amongst three techniques in modelling a series of default risk estimates over the lifetime of each loan, i.e., its term-structure. The behaviour of loans can be described using a nonstationary and time-dependent semi-Markov model, though we model its elements using a multistate regression-based approach. As such, the transition probabilities are explicitly modelled as a function of a rich set of input variables, including macroeconomic and loan-level inputs. Our modelling techniques are deliberately chosen in ascending order of complexity: 1) a Markov chain; 2) beta regression; and 3) multinomial logistic regression. Using residential mortgage data, our results show that each successive model outperforms the previous, likely as a result of greater sophistication. This finding required devising a novel suite of simple model diagnostics, which can itself be reused in assessing sampling representativeness and the performance of other modelling techniques. These contributions surely advance the current practice within banking when conducting multistate modelling. Consequently, we believe that the estimation of loss reserves will be more timeous and accurate under IFRS 9.
    \end{abstract}
     
     
     \JEL{C32, C52, G21.}
    
    \rule{1\columnwidth}{0.4pt}
}

\noindent Word count (excluding front matter and appendix):  10013  

\subsection*{Disclosure of interest and declaration of funding}
\noindent This work is financially supported wholly/in part by the National Research Foundation of South Africa (Grant Number 126885), with no known conflicts of interest that may have influenced the outcome of this work. The authors would like to thank all anonymous referees and editors for their extremely valuable contributions that have substantially improved this work.



\newpage

\section{Introduction}
\label{sec:intro}

Lending poses the fundamental risk of capital loss should the borrower fail to repay their loan, which necessitates the accurate prediction of the borrower's underlying \textit{probability of default} (PD).
This task usually involves finding a statistical relationship between a set of borrower-specific input variables and the binary-valued repayment outcome (i.e., defaulted or not) over some outcome period. The literature on this particular classification task is considerable and spans various forms of supervised statistical learning, including machine learning; see \citet{hand1997statistical}, \citet{siddiqi2005credit}, \citet{thomas2009consumer}, \citet{hao2010review},  \citet{baesens2016credit}, and \citet{louzada2016review}. 
However, these credit rating systems focus mostly on producing a conservative PD-estimate that remains static (but stressed) over the lifetime of each loan, often by design, as discussed by \citet{crook2010dynamic}. The broad goal of such systems is to facilitate the estimation of regulatory and economic capital, which should absorb any catastrophic (or unexpected) losses under the Basel framework from the \citet{basel2019}. 
Any temporal effects that might affect the PD during loan life are therefore largely ignored, together with any macroeconomic influences; particularly since the latter is already assumed to be stressed to a recession-like level during PD-estimation. Doing so renders the resulting PD-estimates as \textit{through-the-cycle} (TTC) in that they should at least approximate the long-run averages of 1-year historical default rates over a full macroeconomic cycle, as required during capital estimation. 
While these TTC PD-estimates are certainly stable over time by design, they are also typically inaccurate within any other setting besides capital estimation.

The alternative, as reviewed and compared by \citet{crook2010dynamic}, is a \textit{point-in-time} (PIT) rating system, which produces more dynamic PD-estimates that agree more closely with the observed variation in default risk over loan life, as well as incorporate any temporal macroeconomic effects. 
Such dynamicity is perhaps inappropriate for capital estimation since capital levels should preferably not fluctuate wildly over time. 
However, these PIT-based PD-estimates are more flexible in that they can be consumed across a greater variety of settings than their TTC-based counterparts. 
In fact, the introduction of the IFRS 9 accounting standard by the \citet{ifrs9_2014} provided additional impetus for such dynamicity in PD-modelling. Under IFRS 9, a financial asset’s value should be comprehensively adjusted according to a bank’s (evolving) expectation of the asset’s \textit{credit risk} over time, i.e., the potential loss induced by default. In principle, the bank willingly forfeits a portion of its income at each reporting period into a loss provision, which should ideally offset any amounts that are written-off in future; thereby smoothing away volatility. The provision size is regularly updated based on a statistical model of the asset's \textit{expected credit loss} (ECL), wherein the PD is embedded as perhaps the most important risk parameter. The bank dynamically adjusts its loss provision either by raising more from earnings or releasing a portion thereof back into the income statement, defined respectively as an \textit{impairment loss} or \textit{gain} according to \S 5.5.8 in IFRS 9. 
Any unnecessary impairment variation, perhaps due to an inaccurate ECL-model, would therefore directly affect a bank's income statement.

The ECL may itself be calculated by following a 3-stage approach (\S 5.5.3, \S 5.5.5) given the extent of the perceived deterioration (or improvement) in credit risk. 
In principle, the ECL-estimate of each impairment stage should become increasingly severe, thereby allowing the timeous -- and more dynamic -- recognition of credit losses; see \S B5.5.3 of IFRS 9, \citet{PwC_2014}, \citet{EY_2018}, \citet{botha2025sicr}, and \citet{BothaVerster2025}.
In achieving such dynamicity, and especially for Stages 1-2, \citet{skoglund2017} noted that risk models need to project default risk ideally over various time horizons across loan life and against the changing macroeconomic background.
This rather non-trivial task implies the estimation of a marginal (or PIT) PD as a function of a rich set of input variables, including macroeconomic covariates. In fact, \citet{JakubikTeleu2025} found that including such covariates can increase the effectiveness of credit risk modelling and stress-testing using `Bayesian Model Averaging', especially during uncertain times of macroeconomic upheaval.
Regardless, these inputs are measured at each discrete period $t=t_1,\dots,\mathcal{T}$ during a loan's lifetime $\mathcal{T}$, starting from its time of initial recognition $t_1$. The collection of these PD-estimates over the lifetime of a loan is then called the \textit{term-structure} of default risk. This term-structure typically manifests as a non-linear and right-skewed curve over loan life.
Put differently, the non-linearity of the term-structure speaks to the required dynamicity of ECL-estimates under IFRS 9.

However, there are certain modelling challenges to rendering such dynamic and time-sensitive PD-estimates. 
Chief among them is due to the fact that `default' is not necessarily an absorbing state into which a loan is forever trapped, as discussed by \citet[pp.~73-83]{botha2021phd}. This dynamicity is acknowledged in both \S36.74 of the Basel framework and in Article 178(5) of the Capital Requirements Regulation, promulgated by the \citet{eu2013CRR} for EU-markets, which requires banks to rate loans as performing whenever default criteria cease to apply. If `default' is structured as a transient state during PD-estimation, then one can leverage the full credit histories that are otherwise etched with multiple cycles of curing from default and defaulting again. 
Another major modelling challenge arises from the fact that `default' is not the only failure-inducing event, despite its importance in credit risk modelling. Other events that may ultimately affect the risk of loss under IFRS 9 include prepayments (or early settlement), write-offs, and restructures. These \textit{competing risks} will preclude the default-event from occurring, as well as affect the size of the risk set over time.
Lastly, default risk is itself a heterogeneous spectrum in that not all loans will have the same PD at the same time point, largely due to differences in the behavioural profiles of borrowers.
If ignored, then all of these factors can inject severe bias into the eventual PD-estimates, thereby attenuating a bank's impairment buffers.

In addressing these modelling challenges, we chiefly contribute an in-depth and empirically-driven comparative study amongst three modelling techniques towards deriving dynamic PD-estimates under IFRS 9. Such a benchmark study does not yet exist explicitly, at least to the best of our knowledge.
While they are themselves well-known, these techniques are selected and applied within a novel context, and improve upon an older technique that is commonly applied, i.e., \textit{Markov-type} models, as reviewed in \autoref{sec:background}.
In ascending order of complexity, and as presented in \autoref{sec:method}, our techniques include: 1) a Markov chain as the baseline model; 2) beta regression (BR); and 3) multinomial logistic regression (MLR). The BR and MLR techniques are applied within a multistate framework whereby the cells of the underlying transition matrix are modelled separately using each technique. 
Doing so has practical relevance in that it addresses the modelling challenges of both a transient default state and of competing risks. Our approach may therefore produce default risk estimates that are less biased than those generated by following other approaches, as reviewed by \citet{BothaVerster2025}.
Within our context, we find the MLR-model to be superior in every regard, followed by the BR-model; both of which supersede the ordinary Markov chain.
By producing more accurate and granular PD term-structures, our results enable managers to meet the core aim of IFRS 9 in recognising credit losses more timeously and accurately. 

Much of the existing literature, as we shall review in \autoref{sec:background}, uses a rather limited set of model diagnostics when evaluating PD-models.
Limited diagnostics would likely not pass muster in practice, especially from the perspectives of external auditors or regulators.
This is especially true given the increasing prevalence of \textit{model risk management} (MRM) as a separate risk type in banking; see the UK-regulator's five MRM-principles in \citet{pra2023PSmodelrisk} and \citet{pra2023SSmodelriskprinciples}. It therefore becomes paramount to benchmark any suite of modelling techniques using a standardised set of more appropriate validation techniques, whose constitution is not always obvious.
In our work, we do not only contribute a comparative study of different modelling techniques, but we shall also contribute a series of more suitable diagnostics for evaluating these models, their predictions, and the underlying data.

Moreover, our work offers a small spectrum of sophistication amongst three techniques that practically allow managers to select an appropriate technique, fully cognizant of the now-apparent trade-off between model complexity \& effort vs accuracy. This choice has practical implications for managing model risk, which is already exacerbated by at least three areas of regulatory constraints of IFRS 9, at least according to \citet{Forrest2024}. These areas include the following. 
Firstly, point-estimates from ECL-models should be unbiased under IFRS 9, which the author argued is not the central aim of statistical modelling, but rather describing and controlling error distributions.
Secondly, IFRS 9 implies a near-constant cycle of adjusting models and their outputs via incremental changes towards attaining unbiased results, which may introduce more problems than it fixes over time; accruing a technical `debt' of sorts.
Thirdly, ECL-calculation under IFRS 9 typically requires a constellation of interconnected models, which increases model risk unavoidably.
In these regards, our work can help navigate these regulatory constraints by delivering a more accurate (less biased) model with greater longevity, which can alleviate the cycles of model adjustments. The more sophisticated MLR-technique demonstrably required but two models, whereas the less sophisticated BR-technique necessitated six models and some assumptions. The implication is that a more sophisticated modelling approach can decrease the size of the model constellation, and hence decrease model risk in practice.

The remainder of this paper's layout is as follows. We briefly describe our data in \autoref{sec:results_calibration}, which spans a richer and more granular collection of time-fixed, time-varying, macroeconomic, and idiosyncratic factors; all of which engender greater model performance.
Having fit our models to South African mortgage data, we test the representativeness of subsampled data using a novel but simplistic method, which can itself be reused in other contexts.
Thereafter, we provide the modelling results in \crefrange{sec:results_markov}{sec:results_models} respective to each technique. We formulate a few model diagnostics in \autoref{sec:comparison} for validating multistate models. Thereafter, the predictions from each model are compared over time and at the portfolio-level. From these predictions, we construct the term-structures of default risk and compare them to the empirical term-structure. The study is then concluded in \autoref{sec:conclusion} by some final remarks. In \autoref{sec:Appendix}, we present ancillary material and review the basics of BR-models, an application of Cook's distance towards identifying outliers, a goodness-of-fit analysis, the fundamentals of MLR-models, and the input space of each fitted model. 
The R-based source code of our work is published in a GitHub-repository, as maintained by \citet{botha2025sourcecode}.
Ultimately, we believe that these contributions surely advance the current practice in multistate PD-modelling towards producing timeous and accurate ECL-estimates under IFRS 9.

\section{A review of Markov-type and multistate models for deriving PD-estimates}
\label{sec:background}

A brief review of Markov-models is given in \autoref{sec:background_Markov}, with a particular focus on their demerits within our context. Thereafter, we review in \autoref{sec:background_beta_MLR} a small variety of alternative modelling techniques in estimating dynamic PD-estimates within a multistate setup. Lastly, the implications of this literature review are discussed in \autoref{sec:background_implications} towards formulating the premises of our study.

\subsection{Markov-models and their drawbacks}
\label{sec:background_Markov}

One class of portfolio-level modelling techniques that can overcome the aforementioned challenges (recurrent events, competing risks, and heterogeneity) is that of \textit{Markov} models. In particular, a dynamic phenomenon (e.g., delinquency) is modelled as a stochastic process that depends only on the current state. 
The simplest such model, a first-order Markov chain, was first explored in the credit domain by \citet{cyert1962} in estimating the size of an accounting allowance for offsetting doubtful debts in future. In their seminal work, the overdue balances of retail store accounts were aged and classified at each period into a set of ordered bins, each of which is progressively more in arrears than the last. As an example, consider the bins: 30 days past due (DPD), 60 DPD, 90 DPD, and 120+ DPD; the last of which serves as an absorbing state that signifies debt write-off. Together with a paid-up/settled state, these delinquency states (or arrears categories) constitute the state space within the transition matrix $T$, thereby incorporating all competing risk events. The authors then used Markov theory in estimating the write-off probability (and its variance) towards setting the allowance; see Appendix A.1 in \citet{botha2021phd} for a worked example.
\citet{corcoran1978use} extended this work by first stratifying the data by loan size before estimating $T_s$ within each stratum $s$, which improved the prediction accuracy. 
\citet{vanKuelen1981note} further refined this work, having corrected the method by which overdue invoices are aged into delinquency states; thus recognising partial payments. 
In addition to these studies, \citet{crook2010dynamic} surveyed a few other works that ultimately moors the use of Markov chains in the modelling of credit risk.

Despite the acclaim of Markov chains, most of the aforementioned studies made two critical assumptions when modelling default risk: 1) that $T$ is largely stationary over time; and 2) that the population is homogeneous regarding payment behaviour. 
However, \citet{frydman1985testing} tested and empirically rejected these assumptions by comparing both non-stationary and stationary Markov chains against an extension thereof -- the \textit{mover-stayer} model. Their results showed that both types of Markov chains can substantially under-predict the observed transition rates, largely due to heterogeneous payment behaviour within certain states of $T$. Conversely, the more accurate mover-stayer model can account for heterogeneity quite simplistically by assuming that only a certain portion of loans (called `movers') can exit each state, whereupon they will move according to another transition matrix. 
Having used data from Standard \& Poor (S\&P), \citet{bluhm2007} demonstrated the poor fit of a time-homogeneous Markov chain in modelling corporate credit rating migrations, with each rating representing a state in $T$. They improved the fit substantially by using an interpolation-based approach, thereby allowing the chains to evolve over time in recognition of the inherent heterogeneity.
\citet{frydman2008twoMarkovChains} further integrated such heterogeneity by building a mixture model from two independent Markov chains, having used the same S\&P rating data. The authors showed that some (similarly-rated) firms will transit at different speeds; i.e., the durations within certain states are not exponentially distributed, which violates a key property of Markov chains. The proposed solution presupposes that there are two latent sub-populations that each move differently across states and, more importantly, have their own distinct migration speeds. Given both its current rating and the history thereof, a firm can alternate probabilistically between either sub-population as it ages, and therefore exhibits vastly different transition probabilities relative to those of a similarly-rated firm with a different rating history.
These studies, alongside the works of \citet{nystrom2006credit}, \citet{Pasricha2017}, and \citet{Chamboko2020} certainly show that the dynamics of default risk are heterogeneous and decidedly non-Markov. 
However, and despite incorporating at least some of the heterogeneity, the resulting models remain at the (sub)portfolio-level, which means that the problem of heterogeneity is only partially solved. We therefore conclude that using a simple Markov chain (or some of its extensions) would likely be inadequate in producing granular PD-estimates that are both accurate and sufficiently dynamic across either calendar time or loan life.

\subsection{Alternatives to Markov-models: Beta regression \& multinomial logistic regression}
\label{sec:background_beta_MLR}

A nonstationary Markov chain implies a time-dependent transition matrix $T(t)$, where each matrix cell $T_{kl}(t)$ represents an element of a broader time series over $t$. In particular, $T_{kl}(t)=\hat{p}_{kl}(t)$ denotes the estimated transition probability $\hat{p}_{kl}(t)$ from state $k$ to $l$ between times $t-1$ and $t$. When modelling such percentage-valued panel data, one can use a class of techniques known as \textit{beta regression} (BR) models from \citet{ferrari2004beta}, which can incorporate any set of input variables. Thus far, BR-models have predominantly been applied in modelling the \textit{loss given default} (LGD) risk parameter, denoted by the random variable $L\in[0,1]$.
For example, \citet{calabrese2010bank} represented $L$ as a discrete-continuous mixture respectively between write-off risk and the loss severity given write-off.
In modelling this mixture using defaulted Italian loans, they applied another nonparametric mixture of two Beta kernel estimators, thereby contending with the known bimodality of $L$.
\citet{Huang2011} proposed a generalized Beta regression framework for modelling systematic risk in both $L$ and the PD, having used a time series of (aggregated) realised LGD-values together with a simulation study.
Based on goodness-of-fit, \citet{Yashkir2013} favourably compared both beta and inflated beta regression models for $L$ against a few contenders, though found that the choice of input variables outweighs that of the modelling method.
\citet{tong2013zero} fit a semi-parametric zero-adjusted gamma model, which outperformed the baseline model: an \textit{ordinary least squares} (OLS) regression model for a beta distributed $L$.
Having used Jordanian corporate loans, \citet{Jaber2020} explored several link functions within beta regression towards modelling the LGD, and found the best fitting function to be the probit.
In following a Bayesian approach, \citet{kiefer2007probability} derived PD-estimates using expert opinions with a beta distribution.
Ultimately, and given its relative popularity, we certainly think it worthwhile to use beta regression in modelling $T_{kl}(t)$ over time as a function of a few input variables; an area hitherto unexplored in literature.

In fully catering for any degree of heterogeneity during PD-modelling, the modus operandi should clearly veer away from directly predicting the portfolio's aggregate behaviour, and rather towards predicting that of its constituent loans; i.e., loan-level modelling. 
To this end, \citet{smith1995forecasting} adopted a nonstationary Markovian structure wherein they developed loan-level forecasting models from loan observations that reside within each cell of $T(t)$ at any point $t$ of their lifetime. Together, these `cell-level' models (or sub-models) form a broader and single \textit{multistate} model, in that a loan's history forms a sample path from a broader stochastic process, while individual models govern the various transition types; see \citet{Chamboko2020}. \citet{smith1995forecasting} then considered two competing regression models, simple linear regression (using OLS) and \textit{multinomial logistic regression} (MLR), in predicting the loan-level transitions from state $k$ to four nominal-valued states: 1-Current and 2-Delinquent; and the absorbing states, 3-Written-off and 4-Settled. Within each cell of $T(t)$, either model predicts the corresponding transition probability $p_{kl}(t, \boldsymbol{x}_i
)$ 
given the characteristics $\boldsymbol{x}_i$ of each loan $i$. The authors built one OLS-model per starting state for a loan transiting to any of the other three states; and one MLR-model that can simultaneously estimate all destination state transitions $l$ for any loan within a particular starting state $k$. 
More importantly, the authors successfully predicted these loan-level state transitions using a fairly rich and varied input space 
within each of their cell-level models in $T$, thereby incorporating heterogeneity.
Their work partly inspires the framework in which we shall conduct our own study of multistate PD-modelling.

Over the next few decades, the seminal work of \citet{smith1995forecasting} has been extended in various ways, thereby practically enshrining the production of multistate PD-estimates as a function of input variables.
\citet{grimshaw2011markov} built stratified binary logistic regression models from observations within only the most crucial of cells of a 7-state transition matrix. Regarding the other cells, the authors suggested a so-called \textit{Empirical Bayesian} estimator for augmenting the mean transition probability $p_{kl}(t)$, such that expert beliefs about overall state volumes can be incorporated.
Similarly, \citet{Arundina2015sukuk} compared a four-state MLR-model against a neural network in predicting credit rating migrations of Sukuk corporate bonds, having used a variety of bond-level input variables and macroeconomic covariates.
Using Indonesian credit card data, \citet{adha2018multinomial} favourably compared a 3-state MLR-model against a parametric spline regression model (using only time until the event as `input') in predicting the hazard of either default or attrition.
Aside from MLR-models, \citet{Gaffney2014} presented a transition-based framework for estimating losses, having used Irish residential mortgage data. Their multistate framework chiefly relies on two \textit{intensity models}\footnote{Although within the biostatistical domain, \citet{putter2007tutorial} explained intensity models as a generalisation of a competing risks Cox regression model from the survival analysis literature.} from \citet{kelly2016good}, which respectively predict the defaulting and curing probabilities as a function of the time spent in each state, macroeconomic covariates, and loan-level inputs.  
Relatedly, \citet{leow2014intensity} built six intensity models within a four-state delinquency-based Markov framework, having used retail credit card data in predicting the various loan-level transition probabilities.
In extending their work, \citet{djeundje2018intensity} used a logit link function within the same intensity models whilst embedding the baseline hazard function using flexible B-splines.
All together, these studies demonstrate the greater extent to which heterogeneity can be embedded via input/predictor variables within multistate loan-level PD-models, at least relative to their portfolio-level counterparts that cannot do so.

\subsection{Implications for the current study}
\label{sec:background_implications}

Each of the aforementioned PD-focused studies has clearly contributed (or improved upon) a loan-level approach to multistate PD-modelling. However, it remains yet unclear how these methods might compare in their performance against one another. Our work can help generate such insights in conducting a benchmark study, which would be of value to the practitioner. The closest attempt at such an endeavour is still the seminal work of \citet{smith1995forecasting}, who explicitly compared OLS-models against MLR-models in producing dynamic PD-estimates. We shall extend their work by comparing MLR-models for lifetime default risk against a few modelling techniques other than simple linear regression.
Moreover, classical Markov chains remain popular in producing simple PD-estimates, despite wrongly assuming stationarity and homogeneity. Choosing a Markov chain over a more sophisticated loan-level technique (e.g., MLR-models) might carry an untenably high opportunity cost, which is itself largely an unstudied problem. We therefore believe that including at least a Markov chain model in the eventual benchmark study would be beneficial, given the model's prevalence within existing literature. 
Furthermore, there appears to be some novelty in using beta regression in modelling transition rates specifically, which is why we include this technique in our benchmark study.

For these reasons, our chosen modelling techniques include: 1) a simple stationary Markov chain (which cannot consume any input variables) that serves as a baseline model; 2) a beta regression model that can only leverage portfolio-level inputs; and 3) a multinomial logistic regression model that incorporates both portfolio-- and loan-level inputs. All of these techniques are used in estimating the nonstationary elements within the time-dependent transition matrix $T(t)$ of an overarching multistate semi-Markov model. These techniques are deliberately chosen in ascending order of complexity, at least based on their ability to consume a wide variety of input variables. In so doing, one should be able to embed the known heterogeneity underlying $T(t)$ to progressively greater extents, thereby addressing one of the weaknesses of Markov models. In fact, we shall demonstrate exactly this phenomenon across our selected modelling techniques in due course.

\section{Three models for deriving lifetime PD-estimates}
\label{sec:method}
We outline a baseline Markov chain in \autoref{sec:Method_MarkovChain} against which the more sophisticated modelling methods will be compared. One such method is a \textit{beta regression} (BR) model that can relate a set of portfolio-level input variables (including macroeconomic covariates) to outcomes between 0 and 1, as discussed in \autoref{sec:Method_betaRegression}. Lastly, we outline in \autoref{sec:Method_MultiLogistic} a \textit{multinomial logistic regression} (MLR) model for predicting a categorical outcome using various loan-level input variables and macroeconomic covariates.

\subsection{A baseline Markov chain for predicting the loan status using a transition matrix \texorpdfstring{$T$}{Lg}}
\label{sec:Method_MarkovChain}

\begin{figure}[ht!]
    \centering\includegraphics[width=0.75\linewidth,height=0.27\textheight]{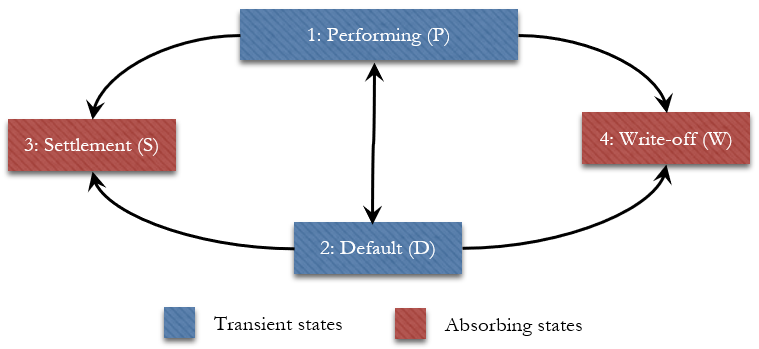}
    \caption{The state space in which loans may reside at any point of their lifetimes.}\label{fig:StateSpace}
\end{figure}

Due to borrower optionality and the vicissitudes of life, a loan may reside in any one of the following four states at any point $t$ of its lifetime $\mathcal{T}$: 1) Performing (P), Defaulted (D), Settled (S), and Written-off (W). A performing loan is typically up-to-date on its payments, though it may accrue payments in arrears until reaching the default threshold, at which point the loan transits to state D. From either of these two transient (and communicating) states $\{\mathrm{P},\mathrm{D} \}$, a loan may also move into one of the two absorbing states, $\{\mathrm{S},\mathrm{W} \}$, whereupon observation of the loan ceases thereafter. Practically, and aside from behavioural profiles, the only difference between S and W is a non-zero outstanding balance for W that will need to be written-off as a credit loss. As illustrated in \autoref{fig:StateSpace}, our state space is intelligently designed to account for both competing risks and recurrent events, particularly since loans may have various cycles of defaulting and curing again in reality.
More formally, and in following \citet[\S 1]{norris1997markov}, let $Y_t\in\mathcal{S}$ denote a random variable that can assume one of these four states at time $t$ in our state space $\mathcal{S}\in\{\mathrm{P},\mathrm{D},\mathrm{S},\mathrm{W} \}$. The sequence $Y_{t_1},\dots, Y_{t_\mathcal{T}}$ then forms a discrete-time first-order Markov chain $(Y_t)_{t\geq 0}$ over discrete-time $t\in\mathbb{Z}_{\geq 0}$. This random process may be estimated from data, particularly since each loan history effectively signifies a \textit{sample path} from the underlying Markov chain.
Assuming stationarity, the transition matrix $T$ that governs this Markov chain will have as entries the transition probabilities $p_{kl}$ from state $k$ to $l$, i.e., $p_{kl}=\mathbb{P}\left(Y_{t}=l \, | \, Y_{t-1}=k \right)$ between any two points in time $t-1$ and $t$. From \citet{anderson1957statistical}, the maximum likelihood estimates of each $p_{kl}$ is $n_{kl}/n_k$, where $n_{kl}$ is the number of observed transitions from $k$ to $l$ across the sampling window, while $n_k$ denotes the number of total transitions starting in $k$. 
The resulting $T$ is therefore expressed as 
\begin{equation} \label{eq:TransMatrix_stationary}
    T=\begin{bmatrix}
    p_{\mathrm{PP}} & p_{\mathrm{PD}} & p_{\mathrm{PS}} & p_{\mathrm{PW}} \\
    p_{\mathrm{DP}} & p_{\mathrm{DD}} &p_{\mathrm{DS}} & p_{\mathrm{DW}} \\
    0&0&1&0\\
    0&0&0&1
    \end{bmatrix} \, .
\end{equation}

Our state space $\mathcal{S}$ is deliberately similar to that of \citet{smith1995forecasting}, thereby promoting comparability and simplicity. By transforming the number of payments in arrears into states, the 7-state model from \citet{grimshaw2011markov} is rather excessive in practice and comes at greater computational cost and sparser data, at least anecdotally so. Moreover, our condensed input space allows for incorporating several delinquency-themed variables in predicting the transition P$\rightarrow$D. In particular, these delinquency-themed variables can capture the dynamics ordinarily associated with movements across the authors' 7-state model. Therefore, including extra delinquency-based states beyond that of D is deemed to be wholly unnecessary in our context. Moreover, we constrain our state space in controlling the modelling effort that would otherwise be required for larger spaces, thereby allowing us to maintain a high standard of quality.
While a Markov chain is easy to estimate and implement, its inherent assumption of homogeneity and stationarity has proven to be untenable in practice, as reviewed in \autoref{sec:background}. Practically, it also lacks the ability to account for input variables, at least directly; and would theoretically need to rely on segmentation towards risk-sensitising the matrix. It is for this reason that we use a Markov chain as a baseline model upon which can be improved by subsequent (and more sophisticated) models.

\subsection{Building a beta regression (BR) model for predicting the elements of \texorpdfstring{$T$}{Lg}}
\label{sec:Method_betaRegression}

\begin{figure}[ht!]
    \centering\includegraphics[width=1\linewidth,height=0.4\textheight]{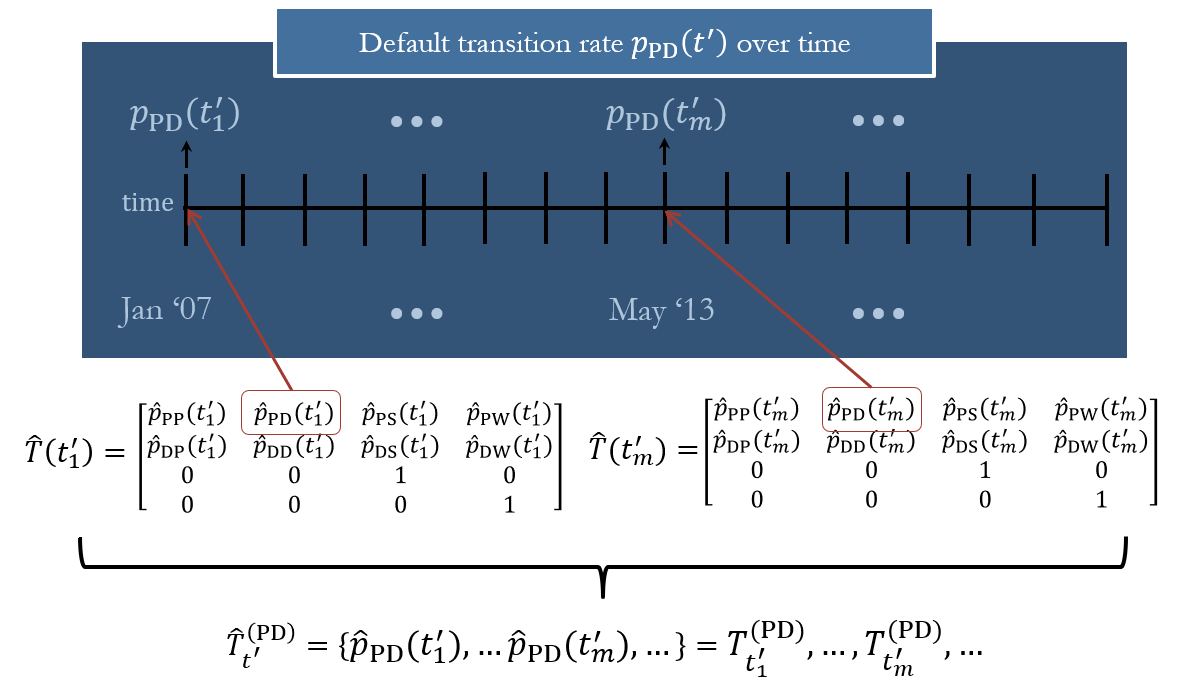}
    \caption{Illustrating the construction process of the outcome variable in BR-modelling for the P$\rightarrow$D transition type, having used the entries within the time-dependent transition matrix $T(t')$.}\label{fig:BR_SampleConstruction}
\end{figure}

The time-homogeneous transition matrix $T$ from \autoref{eq:TransMatrix_stationary} can be easily re-estimated as a time-dependent quantity $T(t')$ over calendar time $t'=t'_1,\dots,t'_n$, e.g., Jan-2007 to Dec-2022. Having partitioned the data by monthly cohort $t'$, each matrix element in $T(t')$ is the time-dependent transition probability $p_{kl}(t')$ from state $k$ to $l$, estimated simply using the same MLE from before as 
\begin{equation} \label{eq:transRate_BR}
    \hat{p}_{kl}(t') = \frac{n_{kl}(t')}{n_k(t')} \, .
\end{equation}
In particular, $n_{kl}(t')$ denotes the number of transitions from $k$ to $l$ during the interval $(t'-1,t']$, while $n_k(t')$ similarly represents the total volume of transitions starting in $k$ during the same interval. For each $(k,l)$-tuple, we assemble the resulting sequence of $\hat{p}_{kl}(t')$-values into a time series $T^{(kl)}_{t'} = T^{(kl)}_{t'_1},\dots,T^{(kl)}_{t'_{n}}$ of specific transition probabilities, as illustrated in \autoref{fig:BR_SampleConstruction}. This time series may then be modelled using a beta regression (BR) model -- itself discussed in \cref{app:betaRegression_basics} -- as a function of portfolio-level input variables, expressed as the variable set $(\boldsymbol{x}_{t'})_{t'\geq t'_1}^{t'_{n}}$ measured over $t'$, where each $\boldsymbol{x}_{t'}=\left\{x_{t'1},\dots, x_{t'p} \right\}$ contains $p$ variables. Put differently, and given the state transition $k$ to $l$ at the $m^{\mathrm{th}}$ time period $t'_m$, the observed pair
\begin{equation} 
    \left(T_{t'_m}^{(kl)}, \boldsymbol{x}_{t'_m}  \right) \nonumber
\end{equation}
therefore constitutes a single observation within the sample from which a BR-model is estimated. In our context, the sample size is roughly $n=190$ such observations, depending on each transition type; which we believe to be adequate for modelling purposes.

Consider the dataset $\mathcal{D}=\left\{t', k, l, T_{t'}^{(kl)}, \boldsymbol{x}_{t'}^{(kl)}, \boldsymbol{z}_{t'}^{(kl)} \right\}$ over calendar/reporting time $t'=t'_1,\dots,t'_{n}$.
We shall construct a set of BR-models with {variable dispersion}, i.e., VDBR-models\footnote{Technically, all our BR-models are VDBR-models, though we have restricted the abbreviation to `BR' for simplicity.}; one such model for each transition type of interest, from state $k$ to $l$. In so doing, we relate the transition rate $T_{t'}^{(kl)}$ over $t'$ with two sets of input variables, denoted by $\boldsymbol{x}_{t'}^{(kl)}$ and $\boldsymbol{z}_{t'}^{(kl)}$, respective to the two components of a BR-model: its mean and its precision. These sets contain predictive information that are specific to the transition type $k\rightarrow l$; thereby embedding heterogeneity at the portfolio-level.
Each BR-model is then constructed by modelling both its mean $\mu_{t'}^{(kl)}$ and precision $\phi_{t'}^{(kl)}$ parameters for state $k$ to $l$, expressed respectively using functions $f_1$ and $f_2$ of the input variables as
\begin{equation} \label{eq:modelForm_BR}
    g_{1}\left(\mu_{t'}^{(kl)}\right) =  f_1\left( \left( \boldsymbol{x}_{t'}^{(kl)} \right)^\mathrm{T} ; \boldsymbol{\beta}^{(kl)} \right) = \eta_{1t'}^{(kl)} \quad \text{and} \quad g_{2}\left(\phi_{t'}^{(kl)}\right) =  f_2\left( \left( \boldsymbol{z}_{t'}^{(kl)} \right)^\mathrm{T} ;\boldsymbol{\theta}^{(kl)} \right) = \eta_{2t'}^{(kl)} \, .
\end{equation}
In \autoref{eq:modelForm_BR}, $g_1(\cdot)$ and $g_2(\cdot)$ are link functions, $\boldsymbol{\beta}^{(kl)}=\left(\beta_1^{(kl)}, \dots, \beta_{p_1}^{(kl)} \right)^\mathrm{T}$ and $\boldsymbol{\theta}^{(kl)}=\left(\theta_1^{(kl)}, \dots, \theta_{p_2}^{(kl)} \right)^\mathrm{T}$ are vectors of estimable regression coefficients, $\eta_{1t'}^{(kl)}$ and $\eta_{2t'}^{(kl)}$ are linear predictors, and both $\boldsymbol{x}_{t'}^{(kl)}=\left\{ x_{t'1}^{(kl)}, \dots, x_{t'p_1}^{(kl)} \right\}$ and $\boldsymbol{z}_{t'}^{(kl)}=\left\{ z_{t'1}^{(kl)}, \dots, z_{t'p_2}^{(kl)} \right\}$ are observations from two different (but possibly overlapping) sets of input variables respective to $\mu$ and $\phi$.

We shall restrict our model specification to the log-log link function for $g_1(\mu)$ and the log link function for $g_2(\phi)$, even though we experiment later with a few other link functions. As such, \autoref{eq:modelForm_BR} becomes
\begin{equation} \label{eq:modelForm_BR_logit}
    g_{1}\left(\mu_{t'}^{(kl)}\right) = \log{\left( - \log \left( \mu_{t'}^{(kl)} \right) \right)} =  \eta_{1t'}^{(kl)}  \quad \text{and} \quad g_{2}\left(\phi_{t'}^{(kl)}\right) = \log{\left( \phi_{t'}^{(kl)} \right)} = \eta_{2t'}^{(kl)} \, ,
\end{equation}
whereafter one can write
\begin{equation} \label{eq:modelForm_BR_Logit_prediction}
    \hat{\mu}_{t'}^{(kl)} = g_1^{-1}\left(\left( \hat{\boldsymbol{\beta}}^{(kl)}\right)^\mathrm{T} \boldsymbol{x}_{t'}^{(kl)}\right) = \tilde{p}_{kl}\left(t', \boldsymbol{x}_{t'}^{(kl)} \right) \, , \nonumber
\end{equation}
which becomes the predicted transition probability $\tilde{p}_{kl}\left(t', \boldsymbol{x}_{t'}^{(kl)} \right)$ from state $k$ to $l$ at $t'$ once the estimates $\left\{ \hat{\boldsymbol{\beta}}^{(kl)}, \hat{\boldsymbol{\theta}}^{(kl)} \right\}$ are obtained; see \autoref{app:betaRegression_basics}.
Lastly, and as used in the linear predictors within \autoref{eq:modelForm_BR_logit}, the input spaces respective to $\mu$ and $\phi$, denoted respectively by $\boldsymbol{x}_{t'}^{(kl)}$ and $\boldsymbol{z}_{t'}^{(kl)}$, consist of portfolio-level variables that are generally described as follows: 
\begin{enumerate}
    \item idiosyncratic variables specific to the particular loan portfolio [p], denoted by $\boldsymbol{x}_{t'[\mathrm{p}]}^{(kl)}$ and $\boldsymbol{z}_{t'[\mathrm{p}]}^{(kl)}$, e.g., the proportion of loans in arrears at $t'$; and 
    \item macroeconomic [m] variables, denoted by $\boldsymbol{x}_{t'[\mathrm{m}]}^{(kl)}$ and $\boldsymbol{z}_{t'[\mathrm{m}]}^{(kl)}$, e.g., the prevailing inflation rate at $t'$.
\end{enumerate}

Practically, we use \autoref{eq:modelForm_BR_logit} in building six different BR-models for the following transition types. Firstly, we build them from the Performing state $k=1$ to each state $l\in\mathcal{S}_\mathrm{P}=\{1,2,3\}$, except for the Write-off state ($l=4$). Secondly, we also build them from the Default state $k=2$ to each state $l\in \mathcal{S}_\mathrm{D} = \{2,3,4\}$, except for the Performing state ($l=1$). The sets $\mathcal{S}_\mathrm{P}$ and $\mathcal{S}_\mathrm{D}$ expediently contain these permissible end states respective to the starting states $k\in\{1,2\}$. The aforementioned exceptions are chosen since the resulting transition rates varied the most over time, which would have needlessly complicated the modelling. We further note that our aggregated data does not include boundary cases of 0/1-valued transition probabilities, as calculated in \autoref{eq:transRate_BR}. Put differently, all proportions are strictly interior within the open interval $(0,1)$, as required by BR-models.

The transition probabilities of the remaining transition types P$\rightarrow$W and D$\rightarrow$P may then be computed by simply subtracting the sum of the other probabilities from one; i.e., 
\begin{equation}
    1-\sum_{u\in \mathcal{S}_\mathrm{P}}{\tilde{p}_{kl}\left(t', \boldsymbol{x}_{t'}^{(kl)} \right)} \quad \text{for } k=1 \quad \text{and} \quad 1-\sum_{u\in \mathcal{S}_\mathrm{D}}{ \tilde{p}_{kl}\left(t', \boldsymbol{x}_{t'}^{(kl)} \right) } \quad  \text{for } k=2 \, . \nonumber
\end{equation}
However, each BR-model will independently output a $\tilde{p}_{kl}\left(t', \boldsymbol{x}_{t'}^{(kl)} \right)$-estimate for $k\rightarrow l$ irrespective of other transition types, which implies that each row sum in the resulting transition matrix may no longer equal one. As a remedy, consider a simple multiplicative scaling approach whereby each final transition probability is obtained as $\acute{p}_{kl}\left(t', \boldsymbol{x}_{t'}^{(kl)} \right)= z\cdot \tilde{p}_{kl}\left(t', \boldsymbol{x}_{t'}^{(kl)} \right)$. This step arises directly from the closure operator of compositional data analysis, as discussed by \citet{aitchison1982tsacod}. This operator is a standard probabilistic regularisation technique that maps any positive vector onto the unit‑sum simplex, whilst preserving all component ratios. Given the state space $\mathcal{S}$, we calculate $z$ by simply solving for it in
\begin{equation}
    z\cdot \left( \sum_{u\in \mathcal{S}}{ \tilde{p}_{kl}\left(t', \boldsymbol{x}_{t'}^{(kl)} \right) } \right) = 1 \quad \implies \quad z = \frac{1}{\sum_{u\in \mathcal{S}}{ \tilde{p}_{kl}\left(t', \boldsymbol{x}_{t'}^{(kl)} \right) }} \, \nonumber ,
\end{equation}
having substituted the missing probability $\tilde{p}_{kl}\left(t', \boldsymbol{x}_{t'}^{(kl)} \right)$ for $k=1$ and $l\notin \mathcal{S}_\mathrm{P}$ with the realised rate $\hat{p}_{kl}(t')$; similarly so for $k=2$. Doing so ensures that each row forms a valid probability distribution within the context of Markov chains, as discussed by \citet{norris1997markov}. Within a prediction setting, one might simply substitute this missing probability with the mean of $\hat{p}_{kl}(t')$ over $t'$, though future work can certainly review this aspect and develop a more appropriate theoretical justification.

As structured within our wider multistate framework, a BR-model can incorporate portfolio-level input variables directly in producing predictions of the (portfolio-level) transition rate $\hat{p}_{kl}(t')$ from \autoref{eq:transRate_BR}. Furthermore, a BR-model can flexibly assume a wide variety of distributional shapes in the outcome variable, as discussed in \autoref{app:betaRegression_basics}, which implies greater accuracy in the eventual predictions. However, a BR-based approach would require six different models in our context, which invariably invites model risk and necessitates greater effort in model development and maintenance. Lastly, a BR-model operates at the portfolio-level and cannot produce account-level estimates. For these reasons, we position this BR-based approach at the middle echelon within our spectrum of sophistication.

\subsection{Estimating the elements of \texorpdfstring{$T$}{Lg} using a multinomial logistic regression (MLR) model}
\label{sec:Method_MultiLogistic}

Recalling the Markov chain $(Y_t)_{t\geq 0}$ from \autoref{sec:Method_MarkovChain}, let $y_{it}$ denote the observed value from $Y_t$ for loan $i$ over its lifetime $\mathcal{T}$ at each discrete time point $t=t_1,\dots,\mathcal{T}$. These $y_{it}\in\{\text{P, D, S, W} \}$ values are nominal in nature and are encoded accordingly for the ending state $l=1,\dots,4$ as
\begin{equation}
    y_{it} = \begin{cases}
        1 \quad \text{if loan $i$ ends in state $l$ = P at time $t$} \\ 
        2 \quad \text{if loan $i$ ends in state $l$ = D at time $t$}\\
        3 \quad \text{if loan $i$ ends in state $l$ = S at time $t$}\\
        4 \quad \text{if loan $i$ ends in state $l$ = W at time $t$}
    \end{cases} \, . \nonumber
\end{equation}

\begin{figure}[ht!]
    \centering\includegraphics[width=1\linewidth,height=0.2\textheight]{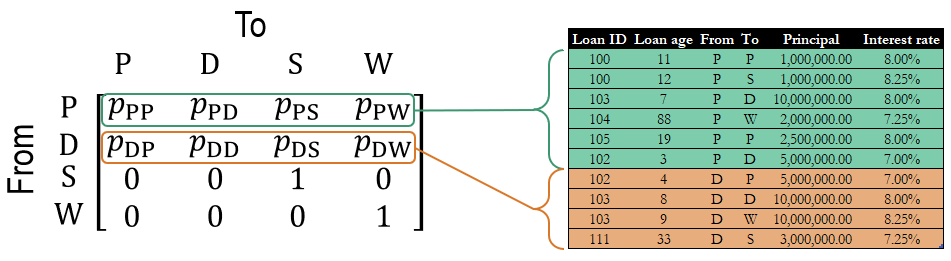}
    \caption{Illustrating the estimation of two MLR-models across various loans within the starting state $k\in\{\text{P, D}\}$, as a function of two covariates, loan amount [\texttt{Principal}] and the central bank policy rate [\texttt{Repo\_rate}].}\label{fig:MLR_Construction}
\end{figure}

As reviewed in \autoref{app:multiLogisReg_basics}, an MLR-model assumes that the ratios of the logarithms of the various transition probabilities $p_{kl}$ can be written as a linear function of input variables $\boldsymbol{x}^{(kl)}$, i.e., $p_{kl}\left(\boldsymbol{x}^{(kl)} \right) = \mathbb{P}\left(Y_{t}=l \, | \, Y_{t-1}=k, \boldsymbol{x}^{(kl)}\right)$. 
In modelling the conditional mean $\mu^{(kl)}_i$ for loan $i$ with a vector of $p$ characteristics $\boldsymbol{x}^{(kl)}_i=\left\{x^{(kl)}_{i1}, \dots, x^{(kl)}_{ip} \right\}$, we shall fit two MLR-models respective to the starting states $k\in\{ \text{P, D} \}$, since the other states are absorbing. These MLR-models are specified using a link function $g(\cdot)$ with a linear predictor $\eta^{(kl)}_i=\beta^{(kl)}_0 + \beta^{(kl)}_1 x^{(kl)}_{i1} + \dots \beta^{(kl)}_p x^{(kl)}_{ip}$, where $\boldsymbol{\beta}^{(kl)} = \left\{ \beta^{(kl)}_0,\beta^{(kl)}_1,\dots, \beta^{(kl)}_p \right\}$ is a vector of estimable regression coefficients. As illustrated in \autoref{fig:MLR_Construction}, the starting state $k$ is itself used as the baseline-category within each MLR-model, which implies the following six model forms (three for each MLR-model), expressed as
\begin{align} 
     \text{Performing (P):} \quad g\left( \mu^{(kl)}_i, \boldsymbol{x}^{(kl)}_i \right) &= \log{\left(\frac{ p_{kl}\left(\boldsymbol{x}^{(kl)}_i\right) }{p_{11}\left(\boldsymbol{x}^{(11)}_i\right)} \right)} = \eta^{(kl)}_i \ \quad \text{for } k=1 \ \text{and } l\in\{ 2,3,4\} \, , \nonumber \\ 
     \text{Default (D):} \quad g\left( \mu^{(kl)}_i, \boldsymbol{x}^{(kl)}_i \right) &= \log{\left(\frac{ p_{kl}\left(\boldsymbol{x}^{(kl)}_i\right) }{p_{22}\left(\boldsymbol{x}^{(22)}_i\right)} \right)} = \eta^{(kl)}_i \ \quad \text{for } k=2 \ \text{and } l\in\{1,3,4\} \, \nonumber.
\end{align}

The regression coefficients $\boldsymbol{\beta}^{(kl)}$ are estimated by maximising the likelihood function, which is achieved using standard numerical procedures, as implemented in the R programming language.
Across all starting states except for S and W, which are absorbing, the predicted (loan-level) transition probabilities $\acute{p}_{kl}(\boldsymbol{x}^{(kl)})$ can then be written for loan $i$ as 
\begin{equation} \label{eq:modelForm_MLR_probs1}
    \acute{p}_{kl}\left(\boldsymbol{x}^{(kl)}_i \right) = \frac{\exp{\left( \eta^{(kl)}_i \right)}}{ 1 + \sum_{j=1}^{4}{\exp{\left( \eta^{(kj)}_i \right) } } } \quad \text{for } l \ne k \quad \text{and} \quad \acute{p}_{kl}\left(\boldsymbol{x}^{(kl)}_i \right) = \frac{1}{ 1 + \sum_{j=1}^{4}{\exp{\left( \eta^{(kj)}_i \right) } } } \quad \text{for } l=k \, .
\end{equation}
Lastly, the input variables $\boldsymbol{x}^{(kl)}$ of the MLR-models reprise those from \autoref{sec:Method_betaRegression}, denoted as $\boldsymbol{x}_{t'[\mathrm{m}]}^{(kl)}$ and $\boldsymbol{x}_{t'[\mathrm{p}]}^{(kl)}$ over calendar time $t'$, but also incorporate the following loan account-level [a] idiosyncratic variables:
\begin{enumerate}
    \item time-fixed variables specific to loan $i$, denoted by $\boldsymbol{x}_{i[\mathrm{a}]}^{(kl)}$, e.g., the chosen payment method;
    \item time-dependent variables specific to loan $i$ and period $t$, denoted by $\boldsymbol{x}_{it[\mathrm{a}]}^{(kl)}$, e.g., the delinquency level.
\end{enumerate}

As for merits, an MLR-model can provide loan-level estimates of the transition rate by incorporating loan-level input variables. Doing so can embed heterogeneity completely and produce even more accurate predictions, which is not possible when using a Markov chain in typical fashion. Furthermore, an MLR-based modelling approach can limit the number of required models: our context required but two such models, as compared to the six BR-models. The implication is that model risk reduces, as well as the amount of effort regarding model development and maintenance. However, an MLR-model assumes that each input variable exerts at least some influence on \textit{each} target state, which may not be true in reality for some states. For example, the policy rate \texttt{[Repo\_rate]} may not be statistically relevant in predicting a transition from the performing state P to the settlement state S, even though it can be relevant in predicting transitions to the other states (default and write-off).

Furthermore, an MLR-model imposes the \textit{Independence of Irrelevant Alternatives} (IIA) assumption, which states that the relative odds between any two transition types (or outcome categories) are unaffected by the presence or characteristics of other transition types; see the Appendix. 
However, we believe that this IIA-assumption is reasonable within our context since the competing destination states are well-separated and correspond to distinct processes. E.g., the underlying (but unobservable) mechanisms driving transitions from default to settlement (D$\rightarrow$S) differ intuitively from those leading to write-off (D$\rightarrow$W). In this example, one such latent factor could be a borrower having excess savings to cover repayments in the event of unemployment. This latent factor will very likely affect D$\rightarrow$S differently than D$\rightarrow$W, and is therefore expected to be largely independent across destination states. This distinctiveness amongst latent factors supports the plausibility of the IIA-assumption.
Furthermore, and even if the IIA-assumption were violated, then we believe that our diagnostics are sufficiently broad in evaluating whether our MLR-models still behave reasonably.

In this section, we introduced and compared three modelling approaches for predicting loan outcomes. A simple stationary Markov chain model was proposed that serves as a baseline, though it cannot incorporate any input variables in its predictions. We then explored a BR-model that can at least leverage portfolio-level inputs (e.g., macroeconomic covariates) in rendering predictions of continuous outcomes $y\in(0,1)$, i.e., transition probabilities. Lastly, we presented an MLR-model that integrates both portfolio- and loan-level inputs in predicting categorical outcomes. Together, these models provide a progressively richer framework for understanding and forecasting loan performance, from which empirical evaluation can follow in the subsequent sections.
\section{Calibrating the various models to South African mortgage data}
\label{sec:results}

In calibrating our modelling techniques from \autoref{sec:method} to data, we discuss the following aspects. In \autoref{sec:results_calibration}, we describe the data and its resampling scheme, along with the process of thematic variable selection. As our baseline model, the estimated transition matrix $\hat{T}$ of the underlying Markov chain is provided and discussed in \autoref{sec:results_markov}. We present in \autoref{sec:results_models} the specifics of both the BR-- and MLR-models in modelling the various transition rates within $\hat{T}$ as functions of input variables. These variables are themselves described in \autoref{app:inputSpace} and span a variety of time-fixed, time-varying, macroeconomic, and idiosyncratic factors.

\subsection{Data calibration: describing the data and resampling scheme}
\label{sec:results_calibration}

We conduct our comparative study using a data-rich portfolio of residential mortgages, as provided by a large South African bank. This longitudinal panel dataset has monthly loan performance observations for each loan $i = 1,...,N$ with $N = 650,715$ 20-year mortgage loans. Each loan $i$ is therefore observed over discrete time $t = t_1,...,\mathcal{T}_i$ from the time of its first month-end observation $t_1$ up to the end of its lifetime $\mathcal{T}_i$. 
These amortising mortgages were sampled from January 2007 up to December 2022, during which time new mortgages were continuously originated, thereby yielding 47,939,860 raw monthly observations of loan repayment performance. Loans that predate the start of this sampling window, i.e., left-truncated loans, are retained along with their subsequent observations throughout this window.
The data and its structure are fully described and documented as comments within the R-based codebase on GitHub, as maintained by \citet{botha2025sourcecode}. Lastly, the selected input variables within the applicable models are explained in \autoref{app:inputSpace}.


Our data is deemed large and we therefore subsample the raw dataset $\mathcal{D}$ into a smaller but still representative sample $\mathcal{D}_S \in \mathcal{D}$. Reasons for doing so include computational expediency, as well as the adverse effect of large sample sizes on $p$-values when testing the statistical significance of regression coefficients; see \citet{lin2013LargeSamples}. Accordingly, we use stratified clustered random sampling by extracting from $\mathcal{D}$ the full credit histories of 200,000 loans, comprising 14,314,925 monthly observations over a period of 192 months. These loan keys are randomly selected within each stratum, where strata are formed by grouping $\mathcal{D}$ based on the date of loan origination, e.g., Jan-2007. Of these 200,000 loans, 70\% are randomly relegated into the training set $\mathcal{D}_T\in\mathcal{D}_S$ whilst the remainder are sorted into the validation set $\mathcal{D}_V\in\mathcal{D}_S$; both of which are used in fitting and evaluating the MLR-models. 
Regarding the resampling scheme for BR-models, we aggregate both $\mathcal{D}_T$ and $\mathcal{D}_V$ to the portfolio-level in calculating the respective transition rates using \autoref{eq:transRate_BR}.

\begin{figure}[ht!]
    \centering\includegraphics[width=0.8\linewidth,height=0.47\textheight]{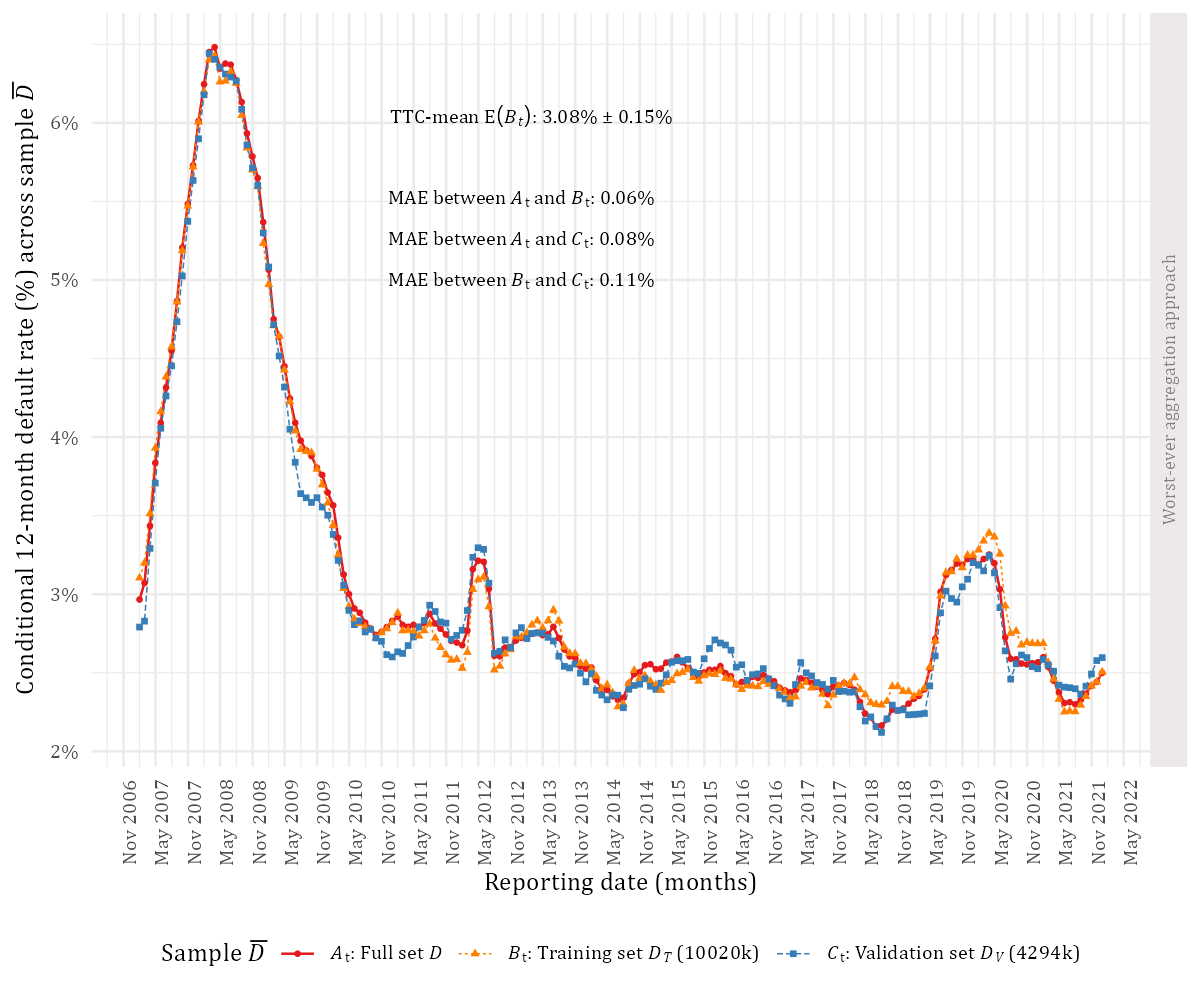}
    \caption{Comparing the 12-month default rates over time across the various datasets. The Mean Absolute Error (MAE) between each sample and the full set $\mathcal{D}$ is overlaid in summarising the line graph discrepancies over time.}\label{fig:DefaultRate_Samples}
\end{figure}

We measure the representativeness of the sets $\left\{ \mathcal{D}_S, \mathcal{D}_T, \mathcal{D}_V, \right\}$ by comparing the $v$-month forward default rate across these sets. Let $D_{it}$ be a Bernoulli random variable that denotes the default status of loan $i$ at time $t$, i.e., 1 if in state D, and 0 otherwise. In creating a $v$-month forward default indicator, we use the \textit{worst-ever} aggregation type from \citet[\S 3.1.3]{botha2021phd} that indicates future default at present time $t$ whenever any of the next $v\geq 1$ statuses $D_{it+1},\dots,D_{it+v}$ equals one. The worst-ever $v$-month conditional probability of a non-defaulted loan $i$ is then $\mathbb{P} \left(\max{\left[D_{it+1},\dots,D_{it+v} \right] = 1} \, | \, D_{it}=0 \right)$. Regarding its estimation, assume that a longitudinal dataset $\mathcal{D}'=\left\{ i, t, d_{it}\right\}$ consists of $d_{it}\in D_{it}$ default status outcomes, whereafter $\mathcal{D}'$ can be partitioned into a series of non-overlapping subsets $\mathcal{D}'(t')$ over calendar time $t'=t'_1, \dots, t'_n$. The aforementioned probability is then estimated at the portfolio-level by the $v$-month default rate, defined at each $t'$ for a given $\mathcal{D}'$ as
\begin{equation} \label{eq:defaultRate_vMonth}
    r\left(t',\mathcal{D}' \right) = \frac{1}{n_{t'}}\sum_{i \, \in \, \mathcal{D}(t')}{\mathbb{I}\left( \max{\left[d_{it+1},\dots,d_{it+v} \right]}=1 \, | \, d_{it}=0  \right)} \quad \text{for } \mathcal{D}'(t')\in \mathcal{D}'\, ,
\end{equation}
where $n_{t'}$ denotes the size of the at-risk population within each subset $\mathcal{D}'(t')$. 
Finally, and in verifying sampling representativeness using \autoref{eq:defaultRate_vMonth}, we graph and compare in \autoref{fig:DefaultRate_Samples} the 12-month default rate over time and across the various datasets. Evidently, the line graphs are reasonably close to one another, with few divergences over time. We furthermore calculate the \textit{mean absolute error} (MAE) between $\mathcal{D}$ and each respective sample, summarised as $\mathcal{D}_T: 0.06\%$ and $\mathcal{D}_V: 0.08\%$; both of which are extremely low per context. Similar results hold for the resampled sets respective to the BR-models. All together, these results suggest that the resampling scheme is indeed representative of the full dataset, which bodes well for training models whose predictions can generalise accurately to the population.

\subsection{Estimating the transition matrix \texorpdfstring{$T$}{Lg} of the Markov chain}
\label{sec:results_markov}

\begin{figure}[ht!]
    \centering
    \begin{subfigure}{0.49\textwidth}
        \caption{Starting from the performing state P}
        \centering\includegraphics[width=1\linewidth,height=0.27\textheight]{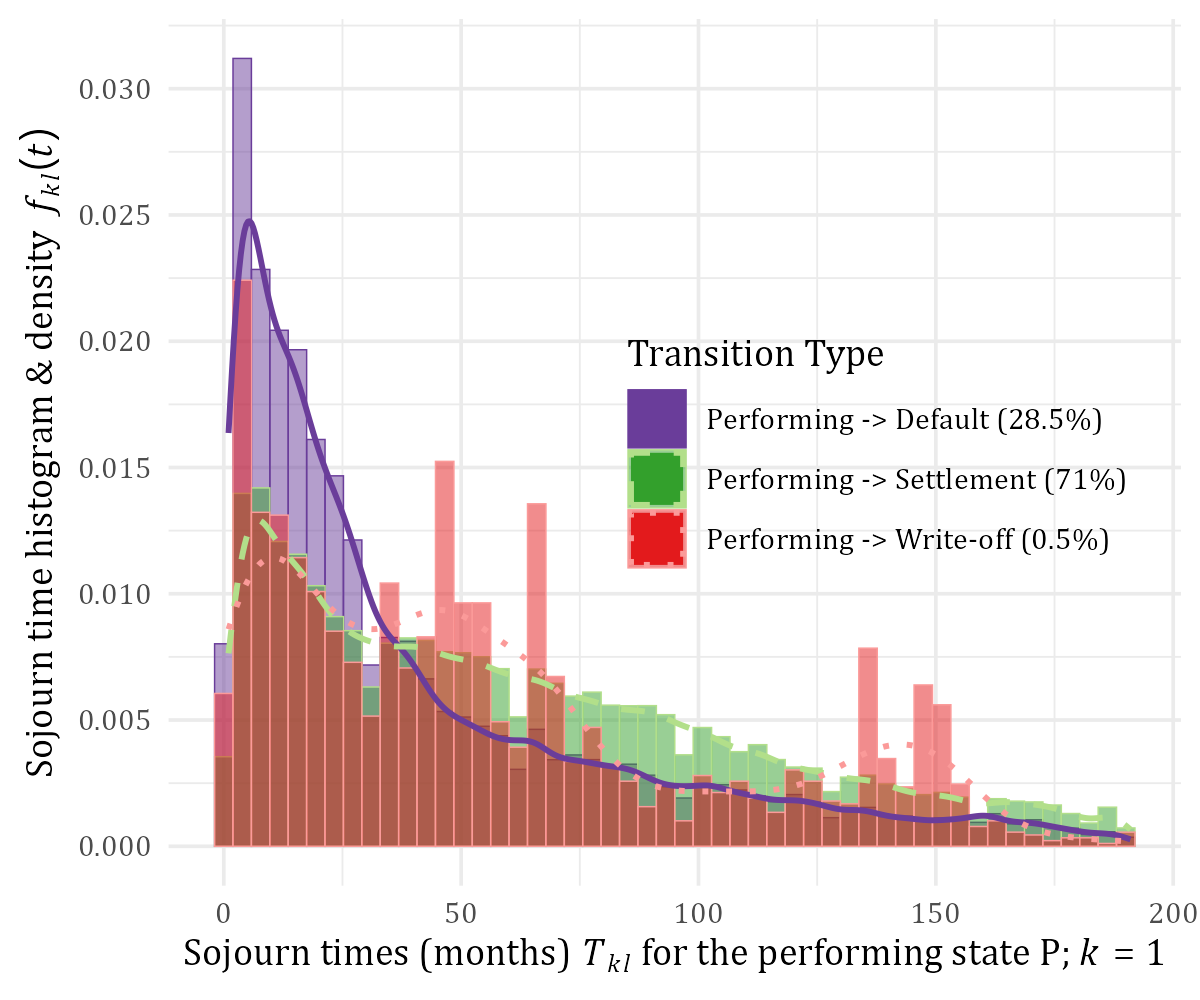}
    \end{subfigure}
    \begin{subfigure}{0.49\textwidth}
        \caption{Starting from the default state D}
        \centering\includegraphics[width=1\linewidth,height=0.27\textheight]{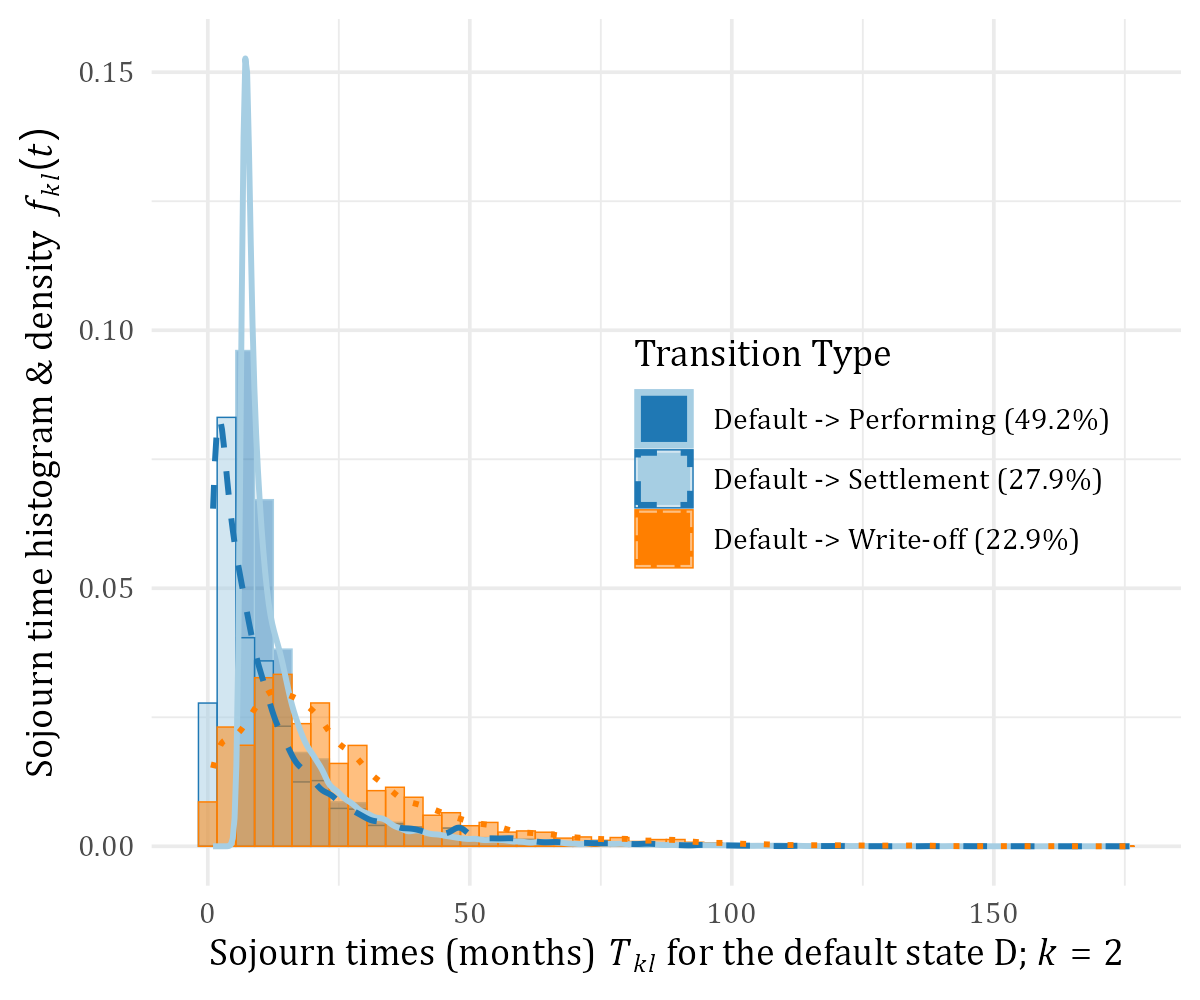}
    \end{subfigure}
    \caption{Histograms and empirical densities of the sojourn times per transition type for the following starting states: performing P in \textbf{(a)}, and default D in \textbf{(b)}. }\label{fig:Hist_SojournTimes}
\end{figure}

Having used the subsample $\mathcal{D}_S$, we obtain an estimate $\hat{T}$ of the time-homogeneous transition matrix, which serves as our baseline model for generating a term-structure of default risk, where $\hat{T}$ is given as
\begin{equation} \label{eq:TransMatrix_stationary_results}
    \hat{T}=\begin{bmatrix}
    \hat{p}_{\mathrm{PP}} &     \hat{p}_{\mathrm{PD}} &     \hat{p}_{\mathrm{PS}} &     \hat{p}_{\mathrm{PW}} \\
        \hat{p}_{\mathrm{DP}} &     \hat{p}_{\mathrm{DD}} &     \hat{p}_{\mathrm{DS}} &     \hat{p}_{\mathrm{DW}} \\
    0 & 0 & 1 & 0\\
    0 & 0 & 0 & 1
    \end{bmatrix} = \begin{bmatrix}
    0.98960 & 0.00297 & 0.00737 & 0.00005 \\
    0.02642 & 0.94634 & 0.01490 & 0.01234 \\
    0&0&1&0\\
    0&0&0&1
    \end{bmatrix} \, .
\end{equation}
Our results suggest that the vast majority of loans remain in their current state over time: 99\% in state P and 94.6\% in state D, presumably due to stellar credit management. These results differ from \citet{Chamboko2020}, who found that only a minority remained performing. In our case, the majority (71\%) of loans that transitioned away from P were settled (S), while most loans (49.2\%) in D that moved away did so back to P; i.e., they are cured from default.
In \autoref{fig:Hist_SojournTimes}, we provide a histogram and empirical densities of the sojourn times $T_{kl}$ per transition type, i.e., the time spent in state $k$ before moving to state $l$. The various histograms are all heavily right-skewed, which is to be expected, though the degree thereof differs markedly; e.g., the distribution of $\text{P}\rightarrow\text{D}$ vs that of $\text{P}\rightarrow\text{S}$. By inspecting these distributions graphically, it is clear that they are surely not exponentially distributed. This result serves as further proof that the Markov-property is indeed violated, given its requirement for sojourn times to be exponentially distributed.

\subsection{Calibrating six BR-models and two MLR-models towards producing \texorpdfstring{$T$}{Lg}-estimates}
\label{sec:results_models}

In fitting either a BR-- or MLR-model, we follow a thematic variable selection process using repeated regressions across themed subsets of input variables. This interactive process is guided by domain expertise, model parsimony, statistical significance, goodness-of-fit (GoF), and model performance on the validation set $\mathcal{D}_V$; some of which are covered in the guidance of \citet{heinze2018variable} regarding variable selection. We use these aspects as `tools' in selecting and scrutinising each input (and its variants) within each final model. 
Firstly, and as explained by \citet{Akaike1998}, a parsimonious model uses the smallest number of inputs relative to the sample size, whilst achieving the maximum GoF-value; itself measurable using the well-known \textit{Akaike Information Criterion} (AIC).
Another useful GoF-measure is the pseudo coefficient of determination for BR-models, denoted as $R^2_{\mathrm{F}}\in [0,1]$, which \citet{ferrari2004beta} described as the squared sample correlation between the linear predictors $\boldsymbol{\eta}^{(kl)}$ and the link-transformed outcome $g\left( \mu^{(kl)}\right)$ for transition type $k\rightarrow l$. For the MLR-models, we use the McFadden pseudo $R^2_\mathrm{McF}\in[0,1]$ measure, which compares the deviance of a fitted model to that of the null model, as detailed by \citet{mcfadden1972conditional} and \citet{menard2000coefficients}.
Secondly, we test the statistical significance of a particular input, which is based on both the Wald-statistic for BR-models (given their small sample sizes) and the likelihood ratio test for MLR-models (given their large sample sizes), as discussed by \citet{heinze2018variable}. 
Thirdly, model performance is evaluated out-of-sample by comparing the model's predictions within $\mathcal{D}_V$ against the observed outcomes, where the exact method depends on the modelling technique; see \autoref{sec:comparison}.
Finally, and in producing insight, these tools are used in evaluating various intermediary models that contain certain subsets of inputs, which are grouped by a thematic and central question of interest; e.g., "\textit{which lagged version of the interest rate is `best' in predicting the outcome?}". The insights are then collected across themes in forming a combined input space, whereafter a stepwise forward selection procedure from \citet[\S 6]{james2013introduction} is run, whereupon the results are finally curated by domain expertise. 
Our thematic selection process is further detailed in the codebase, created by \citet{botha2025sourcecode}. This process culminates in a unique input space for each type of modelling technique and transition, as summarised in \autoref{app:inputSpace}.

In fitting BR-models specifically, we experimented with including vs excluding variable dispersion within the models; i.e., modelling or fixing the precision parameter $\phi$. Although not shown here, our experimental results suggest that a better fit is achieved by explicitly modelling $\phi$, and so the following BR-models are therefore technically VDBR-models. We also experimented with different link functions and, though the difference in results was inscrutably small, we opted for the log-log link function for $\mu$ across all BR-models.
Having finalised the input space, we tested for influential observations $i$ using Cook's distance $D_\mathrm{C}(i)$, as detailed in \autoref{app:cookDistance}. A few observations (1-3) are indeed considered as influential within each BR-model, and are subsequently removed from the sample; thereby slightly improving the overall fit of each model. The final fit statistics are themselves provided in \autoref{tab:BR_model_fitStats}, after refitting the BR-models.
Evidently, both the AIC and the $R^2_\mathrm{F}$ in \autoref{tab:BR_model_fitStats} have broad agreement in the quality of model fit, showing that the BR-model for P$\rightarrow$D has the best overall goodness-of-fit.
Finally, we analysed the Pearson residuals of each BR-model in gauging the model fit, and tested the residual distributions for normality using the \textit{Kolmogorov-Smirnov} (KS) test at a significance level of $\alpha=5\%$. While \citet{ferrari2004beta} acknowledged that the distributional form of Pearson residuals is not exactly known, we thought it prudent to assume that these residuals follow a normal distribution, given its prevalence in linear models. As summarised in \autoref{tab:BR_model_fitStats}, our normality tests show that the residual distributions are indeed approximately normally distributed (or at least symmetrical), with details of this exercise provided in \autoref{app:pearsonResiduals}. All together, these results suggest that all BR-models fit the training data $\mathcal{D}_T$ quite well.

\begin{table}[!ht]
\centering
\caption{Various fit statistics of the final BR-models across transition types, having deleted influential observations. Information criteria include the \textit{Akaike Information Criterion} (AIC). The $R_\mathrm{F}^2$ refers to the pseudo coefficient of determination. Skewness refers to the usual Fisher-Pearson skewness coefficient in summarising the distribution of Pearson residuals of each BR-model. The $p$-values are those originating from a KS-test in testing the residual distribution for normality.}
\label{tab:BR_model_fitStats}
\begin{tabular}{@{}llllll@{}}
\toprule
\multirow{2}{*}{\textbf{BR-model $kl$}} & \multicolumn{5}{c}{\textbf{Fit statistics}} \\
 & Sample size & AIC & $R^2_{\mathrm{F}}$ & Skewness & KS $p$-values \\ \midrule
PP & 190 & -2,039 & 69.96\% & 0.476 & 13.56\% \\
PD & 189 & -2,445 & 87.21\% & 0.710 & 16.12\% \\
PS & 189 & -2,031 & 60.08\% & -0.849 & 3.24\% \\
DD & 190 & -1,269 & 33.27\% & -0.198 & 91.67\% \\
DS & 190 & -1,535 & 63.47\% & 1.405 & 10.67\% \\
DW & 190 & -1,529 & 39.74\% & 0.872 & 38.84\% \\ \bottomrule
\end{tabular}
\end{table}

Regarding the MLR-models, the stepwise forward selection procedure is run on a super-sampled (and smaller) subset of only 50,000 loans that were randomly selected from $\mathcal{D}_T$. In so doing, we shortened the excessively long compute times of the procedure by multiple hours. The MLR-models are however refit on $\mathcal{D}_T$ with the selected variables. During this final training step, we also apply a few natural regression splines on some inputs using expert judgement, which significantly improved the overall model fit; see \autoref{app:inputSpace} for details. The final fit statistics are provided in \autoref{tab:MLR_model_fitStats} for each MLR-model, denoted as P$l$ and D$l$ in predicting the transition types from either P or D to any state $l$. We note immediately that the AIC-measure is not directly comparable, since the sample size changes drastically between the P$l$ and D$l$ MLR-models. Following the rules of thumb from \citet{mcfadden1972conditional} for interpreting the $R^2_\mathrm{McF}$-measure, it would appear that both MLR-models have a strong fit since $R^2_\mathrm{McF}\in [0.2,0.4]$.
Thereafter, the discriminatory power of these MLR-models is assessed using a \textit{receiver operating characteristic} (ROC) curve, as outlined by \citet{fawcett2006introduction}. In conducting ROC-analyses, the multinomial prediction task is first transformed into a series of binary classification tasks, respective to each transition type using indicator functions. 
The resulting ROC-analyses are summarised into the well-known AUC-statistic $\in[0.5,1]$, which is shown in \autoref{tab:MLR_model_fitStats} accordingly.
Evidently, the vast majority of predicted transitions $k\rightarrow l$ have at least a decent level of discriminatory power ($\text{AUC}\geq75\%$). Moreover, six transition types have excellent results ($\text{AUC}\geq90\%$), including the strategically important P$\rightarrow$D type.

\begin{table}[!ht]
\centering
\caption{Various fit statistics of the MLR-models across transition types. The $R^2_\mathrm{McF}$-measure is the McFadden pseudo coefficient of determination. In summarising an ROC-analysis on each transition type $k\rightarrow l$, the AUC is calculated both in-sample within $\mathcal{D}_T$ and out-of-sample within $\mathcal{D}_V$. Each AUC-statistic is accompanied by 95\% confidence intervals, calculated using the DeLong-method from \citet{delong1988comparing}.}
\label{tab:MLR_model_fitStats}
\begin{tabular}{@{}lllllll@{}}
\toprule
\multirow{2}{*}{\textbf{MLR-model} $kl$} & \multicolumn{6}{c}{\textbf{Fit statistics}} \\
 & AIC & $R_\mathrm{McF}^2$ & To state $l$ & Sample size & AUC: $\mathcal{D}_T$ & AUC: $\mathcal{D}_V$ \\ \midrule
\multirow{4}{*}{P$l$} & \multirow{4}{*}{853,356} & \multirow{4}{*}{26.95\%} & P & \footnotesize{9,020,554} & \footnotesize{81.66\% ± 0.141\%}  & \footnotesize{81.62\% ± 0.204\%} \\
 &  &  & D & \footnotesize{27,184} & \footnotesize{98.20\% ± 0.096\%} & \footnotesize{98.23\% ± 0.136\%} \\
 &  &  & S & \footnotesize{66,643} & \footnotesize{75.95\% ± 0.169\%} & \footnotesize{76.10\% ± 0.243\%} \\
 &  &  & W & \footnotesize{434} & \footnotesize{93.42\% ± 1.032\%} & \footnotesize{93.77\% ± 1.381\%} \\
\multirow{4}{*}{D$l$} & \multirow{4}{*}{183,464} & \multirow{4}{*}{28.43\%} & P & \footnotesize{457,448} & \footnotesize{96.24\% ± 0.082\%} & \footnotesize{96.33\% ± 0.119\%} \\
 &  &  & D & \footnotesize{12,810} & \footnotesize{78.23\% ± 0.328\%} & \footnotesize{77.94\% ± 0.485\%} \\
 &  &  & S & \footnotesize{7,108} & \footnotesize{74.87\% ± 0.570\%}& \footnotesize{73.74\% ± 0.840\%} \\
 &  &  & W & \footnotesize{6,000} & \footnotesize{79.64\% ± 0.549\%}& \footnotesize{78.23\% ± 0.862\%} \\ \bottomrule
\end{tabular}
\end{table}

In summary, we calibrated the modelling techniques that were previously introduced in \autoref{sec:method} to South African mortgage data. The dataset was briefly described along with its resampling methodology, followed by affirming the representativeness of the latter using the 12-month default rate. 
The baseline Markov chain model was then estimated and discussed, which offers a foundational view of transition dynamics over all time periods. We then modelled the transition rates as a function of time as well as of portfolio-level input variables using six BR-models. As a more sophisticated alternative, we further modelled account-level transitions over time using two MLR-models with a rich input space. For both classes of models (BR and MLR), we provided various fit statistics and performance measures that suggest our models to be of high quality. Lastly, and as explained, our fitting process (thematic variable selection) is quite broad and relies on a variety of aspects in selecting a highly predictive set of inputs amongst time-fixed, time-varying, macroeconomic, and idiosyncratic factors. It is from this basis that our discussion now veers into model comparison in the following section.

\section{Comparing different multistate models across various diagnostics}
\label{sec:comparison}

Our models cannot be compared directly, though may be compared by aggregating their predictions to the portfolio-level using the 1-month state transition rate of type $k\rightarrow l$. The BR-models already produce predicted transition rates $\tilde{p}_{kl}\left(t', \boldsymbol{x}_{t'}^{(kl)} \right)$ at the portfolio-level over calendar time $t'=t'_1, \dots, t'_n$ and given portfolio-level input variables $\boldsymbol{x}_{t'}^{(kl)}$. For comparison purposes, we therefore need only contend with aggregating those predictions from the loan-level MLR-models to the portfolio-level. Accordingly, the arithmetic average is taken across the loan-level predictions $\acute{p}_{kl}\left( \boldsymbol{x}^{(kl)}_i \right)$ for all loans $i$ in each monthly cohort $\mathcal{D}_V(t')\in\mathcal{D}_V$, all of which reconstitutes the validation set $\mathcal{D}_V=\cup_{t'}{\mathcal{D}_V(t')}$ over $t'$. We then express the MLR-variant of the portfolio-level predicted transition rate at any given $t'$ as 
\begin{equation} \label{eq:aggrTransRate_MLR}
    \tilde{p}_{kl}^{\mathrm{M}}(t') = \frac{1}{n_{t'}}\sum_{i\, \in \, \mathcal{D}_V(t')}{\acute{p}_{kl}\left( \boldsymbol{x}^{(kl)}_i \right)} \quad \text{for } \mathcal{D}_V(t')\in\mathcal{D}_V \,
\end{equation}
where $n_{t'}$ is the size of the subset $\mathcal{D}_V(t')$. For any $k\rightarrow l$, the series of both of these expected estimates $\tilde{p}_{kl}\left(t', \boldsymbol{x}_{t'}^{(kl)} \right)$ from the BR-models and $\tilde{p}_{kl}^{\mathrm{M}}(t')$ from \autoref{eq:aggrTransRate_MLR} can now be compared over $t'$ to the actual (or empirical) time-dependent transition rate $\hat{p}_{kl}(t')$. Note that the corresponding estimate $\hat{p}_{kl}$ from \autoref{eq:TransMatrix_stationary_results} that arises from the Markov chain remains the same across all $t'$.

\begin{figure}[ht!]
    \centering\includegraphics[width=0.75\linewidth,height=0.41\textheight]{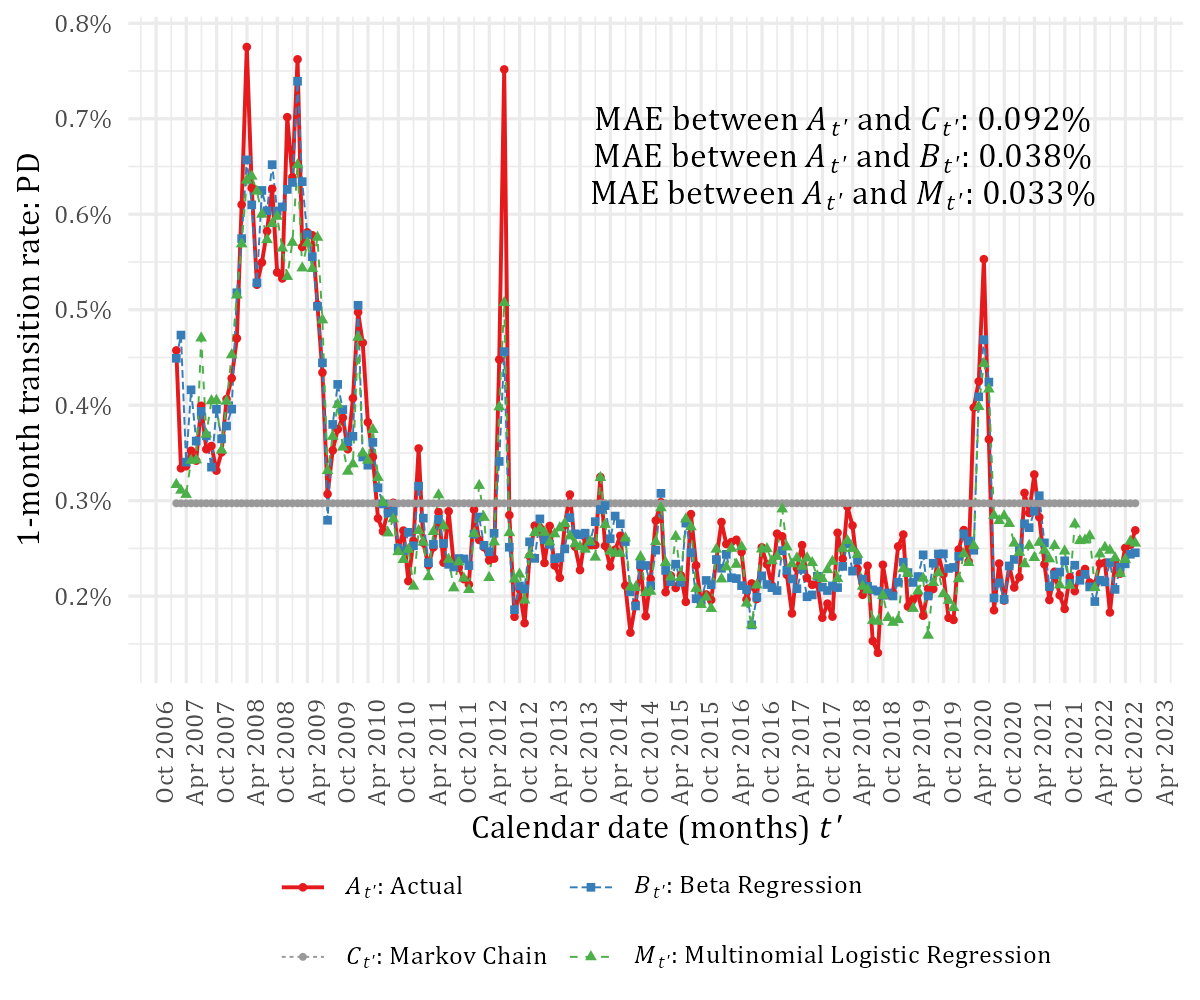}
    \caption{Time graphs of actual vs expected 1-month transition rates for P$\rightarrow$D across various techniques, having used $\mathcal{D}_V$ respective to each technique. The MAE-based AD-statistic from \autoref{eq:AD_stat} is calculated between each actual and expected rate pair in summarising the discrepancies over time.}\label{fig:TransRate_PD}
\end{figure}

\begin{figure}[ht!]
    \centering\includegraphics[width=0.75\linewidth,height=0.41\textheight]{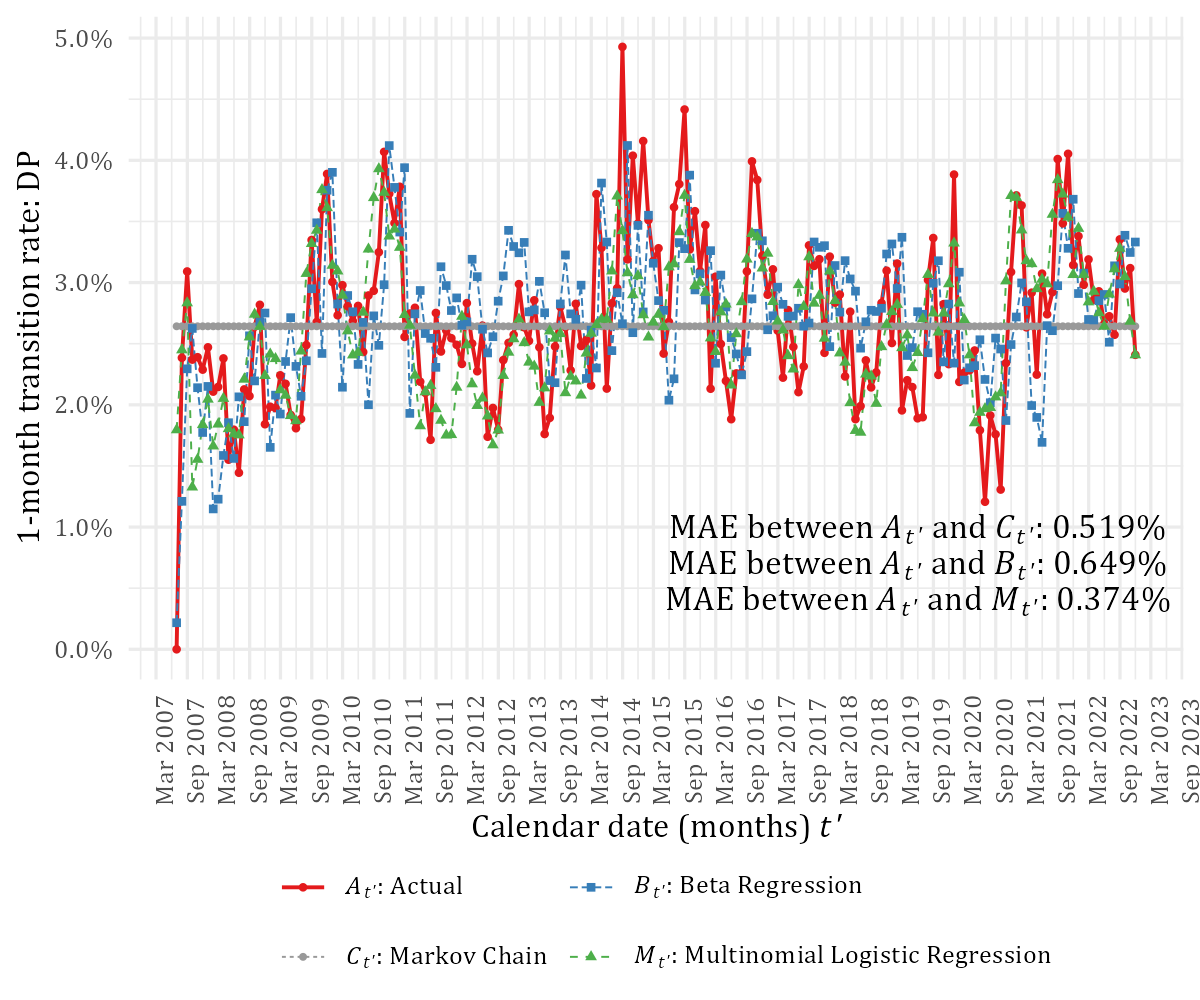}
    \caption{Time graphs of actual vs expected 1-month transition rates for D$\rightarrow$P across various techniques, having used $\mathcal{D}_V$ respective to each technique. Graph design follows that of \autoref{fig:TransRate_PD}. }\label{fig:TransRate_DP}
\end{figure}

We present time graphs in \crefrange{fig:TransRate_PD}{fig:TransRate_DP} of the actual vs expected rates for a specific transition type $k\rightarrow l$. While these graphs include only P$\rightarrow$D and D$\rightarrow$P in the interest of brevity, the time graphs of the remaining types are contained within the codebase, as maintained by \citet{botha2025sourcecode}.
The level of agreement between any combination of actual and expected transition rates can be measured as follows. We compute the MAE between any two rates and compare the subsequent MAE-based values across modelling techniques, which is similar to the use of MAE in testing the sampling representativeness in \autoref{sec:results_calibration}. More formally, and for any pair of rates $\left\{ r_1(t'), r_2(t') \right\}$ observed over calendar time $t'$, we define the MAE-based \textit{average discrepancy} (AD) statistic as 
\begin{equation} \label{eq:AD_stat}
    \text{AD: } \quad \bar{r}_\mathrm{AD}( r_1, r_2 ) = \frac{1}{t'_n - t'_1}\sum_{t'}{\left| r_1(t') - r_2(t') \right|} \, ,
\end{equation}
where $r_1(t') = \hat{p}_{kl}(t')$ is the actual transition rate and $r_2(t')\in \left\{\hat{p}_{kl}, \tilde{p}_{kl}\left(t', \boldsymbol{x}_{t'}^{(kl)} \right), \tilde{p}_{kl}^{\mathrm{M}}(t')  \right\}$ represents any of the expected varieties.
Using the validation sets $\mathcal{D}_V$ respective to each technique, we annotate the AD-statistics in \crefrange{fig:TransRate_PD}{fig:TransRate_DP} respectively for P$\rightarrow$D and D$\rightarrow$P. The results confirm a visual analysis in that the MLR-models agree the closest with reality since they have the smallest $\bar{r}_\mathrm{AD}( r_1, r_2 )$-values. Similar results hold for all other transition types, as summarised in \autoref{tab:MAEs} using the AD-statistic.
Despite being second-rated, the BR-model still outperformed the Markov chain (MC) to a significant degree in most cases. In fact, one can gauge this improvement over the MC-model by expressing the AD-statistic of each competing model relative to that of the MC-model, and subtracting this ratio from 1, whereafter the arithmetic mean is taken. Consequently, the BR-model improved the AD-statistic on average by 58.8\%, whilst the MLR-model improved it even more by 64.1\% on average.

\begin{table}[!ht]
\centering
\caption{MAE-based AD-statistics between actual and expected transition rates across the various transition types $k\rightarrow l$, as produced by the various multistate models. Underlined values indicate the best-in-class performance of a particular modelling technique for a given $k\rightarrow l$.}
\label{tab:MAEs}
\begin{tabular}{@{}lccccc@{}}
\toprule
\multirow{2}{*}{Model} & \multicolumn{1}{l}{} & \multicolumn{4}{c}{To state $l$} \\
 & \multicolumn{1}{l}{From state $k$} & \multicolumn{1}{l}{P} & \multicolumn{1}{l}{D} & \multicolumn{1}{l}{S} & \multicolumn{1}{l}{W} \\ \midrule
\multirow[c]{2}{*}{MC} & P & \footnotesize{0.157\%} &  \footnotesize{0.092\%} & \footnotesize{0.134\%} & \footnotesize{0.004\%} \\
 & D & \footnotesize{0.519\%} & \footnotesize{0.767\%} & \footnotesize{0.516\%} & \footnotesize{0.517\%}  \\
\multirow[c]{2}{*}{BR} & P & \footnotesize{0.104\%} &  \footnotesize{0.038\%} & \footnotesize{0.095\%} & \footnotesize{0.004\%} \\
 & D & \footnotesize{0.651\%} & \footnotesize{0.822\%} & \footnotesize{0.401\%} & \footnotesize{0.366\%} \\
\multirow[c]{2}{*}{MLR} & P & \footnotesize{\underline{0.086\%}} & \footnotesize{\underline{0.033\%}} & \footnotesize{\underline{0.088\%}} & \footnotesize{\underline{0.003\%}} \\
 & D & \footnotesize{\underline{0.374\%}} & \footnotesize{\underline{0.578\%}} & \footnotesize{\underline{0.318\%}} & \footnotesize{\underline{0.331\%}} \\ \bottomrule
\end{tabular}
\end{table}

\begin{figure}[ht!]
    \centering\includegraphics[width=0.8\linewidth,height=0.45\textheight]{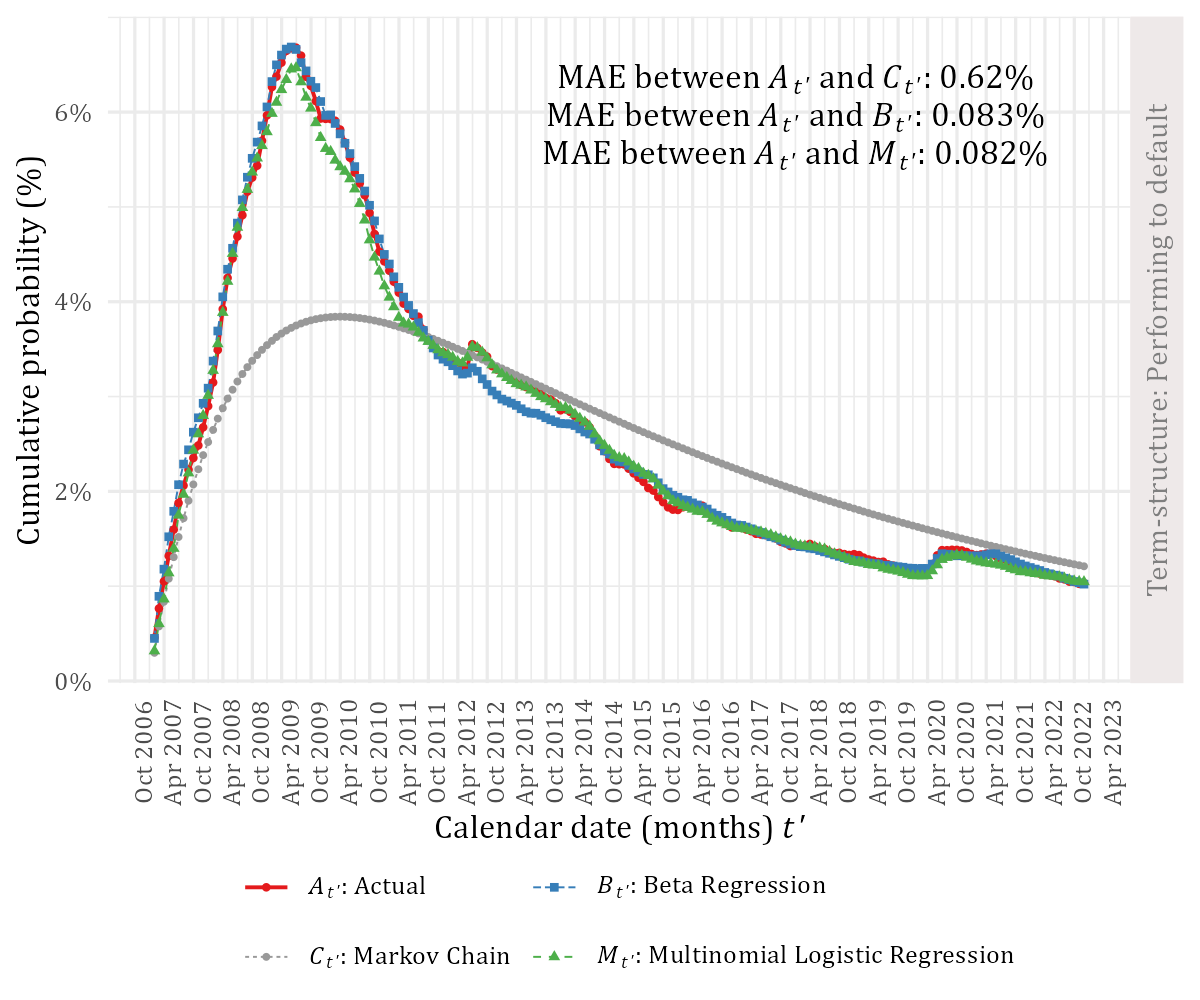}
    \caption{The implied term-structure of actual vs expected transition rates of type P$\rightarrow$D over calendar time $t'$, shown across various techniques. The MAE summarises the discrepancies between each pair of actual and expected rates over time.}\label{fig:TermStructure_estimated}
\end{figure}

Aside from the 1-month state transition rate, we also assess our models by calculating and comparing the implied PD term-structures of all loans within a particular loan cohort $t'=\text{Jan-2007}$, which is observed until maturity. In particular, consider the empirical transition matrices $\hat{T}(t'_1),\dots,\hat{T}(t'_n)$, with elements $\hat{p}_{kl}(t')$ from \autoref{eq:transRate_BR}. These matrices are multiplied recursively as $\acute{T}(t')=\acute{T}(t'-1)\hat{T}(t')$ whilst setting $\acute{T}(t'_1)=\hat{T}(t'_1)$ at the starting period $t'_1$. In so doing, we produce the cumulative transition matrix $\acute{T}(t'_j)$ over the interval $[t'_1,t'_j]$ for some endpoint $j=1,\dots,n$. The elements within $\acute{T}(t'_j), j=1,\dots,n$ represent the actual term-structure of type $k\rightarrow l$ over $t'$; i.e., it is the collection of PIT-probabilities at each $t'_j$, having survived up to $t'_j-1$. In this study, we are however interested only in those cumulative transition rates of type P$\rightarrow$D, which signify the PD term-structure, as shown in \autoref{fig:TermStructure_estimated}. Furthermore, we form a time-dependent expected transition matrix $T^{(\mathrm{e})}(t')$ from the portfolio-level estimates of $p_{kl}(t'), t'=t'_1,\dots,t'_n$ that result from each modelling technique, having aggregated the loan-level predictions from the MLR-model using \autoref{eq:aggrTransRate_MLR}. This expected transition matrix is cumulated similarly over $[t'_1,t'_j]$ until reaching the same endpoint $j$, which can then be compared with $\acute{T}(t'_j)$.
Once again, we show only the P$\rightarrow$D type in \autoref{fig:TermStructure_estimated}. It is quite clear that the Markov chain underestimates the actual term-structure, whereas both the BR-- and MLR-models mirror reality remarkably well, as evidenced by their respective MAE-values (or AD-statistics).



Overall, we studied the predictions from each model in this section, having aggregated these predictions to the portfolio-level over time. Deviations from reality were measured using the average discrepancy (AD). This MAE-based statistic showed the outperformance of the MLR-models over the BR-models; and likewise for the latter over the Markov chain to a significant degree. These diagnostics underscore the relative strengths and limitations of each model class, particularly in their abilities to capture transition dynamics and to respond to macroeconomic externalities. By implication, our results offer a clearer understanding of model behaviour, and avail a small menu of well-tested techniques across the complexity spectrum to the practitioner.

\section{Conclusion}
\label{sec:conclusion}

Only a few works have hitherto studied the problem of loan-level multistate PD-modelling towards producing dynamic PD-estimates over the lifetime of each loan. While the use of Markov-type models remains popular in PD-modelling, it was unclear how they might compare to regression-type models within a multistate setting, or how exactly one might even compare such disparate models.
We filled this gap in literature by contributing an in-depth and empirically-driven comparative study across three modelling techniques of ascending complexity. Each technique was fit on a rich dataset of residential mortgages that was provided by a large South African bank. Firstly, a simple time-homogeneous and stationary Markov chain was fit as a baseline transition-type model against which other techniques may be compared. Secondly, we trained beta regression (BR) models in predicting the time-dependent transition probability as a function of portfolio-level variables, including macroeconomic covariates; all of which appears to be a novel use of BR. Thirdly, multinomial logistic regression (MLR) models were fit in predicting simultaneously the loan-level probability of each type of state transition. Moreover, we used a large and diverse set of input variables when fitting MLR-models, thereby providing insight on the kind of variables that drive certain transition types.

Before comparing the modelling results, we proposed and implemented a simple way of testing the degree to which the sampled data represents the population. In this regard, the $v$-month forward default rate $r\left(t,' \mathcal{D}'\right)$ can be calculated over calendar time $t'$ for any dataset $\mathcal{D}'$. Each dataset will have its own rate series, whereafter discrepancies between any two such rates can be summarised using the mean absolute error (MAE). Smaller values in this MAE-based statistic indicate greater representativeness, and vice versa.
Furthermore, we crafted a method for facilitating a direct and standardised comparison amongst modelling techniques by aggregating their predictions to the portfolio-level. This method of comparison relies again on the aforementioned MAE-based statistic, whereby the 1-month transition rates -- as produced by each modelling technique -- are calculated, compared, and summarised over time into a single portfolio-level statistic. We computed this MAE-based and self-styled \textit{average discrepancy} (AD) statistic for each modelling technique. These AD-statistics confirmed visual analyses in that the performance of the MLR-models are superior to that of all other models across all transition types. Conversely, the Markov chain underperformed in relation to both BR-- and MLR-models, as expected. In fact, both BR-- and MLR-models improved upon the AD-statistic of the Markov chain respectively by 58.8\% and 64.1\%. We obtained similar results when comparing the PD term-structure of transition rates over calendar time; i.e.,  both BR-- and MLR-models drastically outperformed the Markov chain. That said, both of these models performed similarly to each other, which suggests opting for the simpler BR-model, though only in this instance.
Overall, our results demonstrably underscore the value of using more sophisticated regression techniques when estimating the term-structure of default risk, at least relative to using a Markov chain with its resulting model risk.

Despite the contributions of our study, we note the following limitations. While our MLR-models performed admirably amongst other techniques, they impose the same input space across all types of state transitions for a given starting state. Instead of MLR-models, future work can pursue fitting a binary logistic regression model for each transition type, which would tailor the input space accordingly, as in \citet{grimshaw2011markov}. Such models can then be similarly compared to other modelling techniques, including any binary classifier, which lacked from \citet{grimshaw2011markov}. Those classifiers that produce discrete output (e.g., neural networks) can still be considered, given that their predictions can be aggregated to the portfolio-level and evaluated using the AD-statistic.
Keeping MLR as a modelling technique is also worthwhile, though future work can certainly refine its use, e.g., the use of different types of splines and associated knots can be explored and compared in modelling non-linear effects on the transition probabilities.
Regarding BR-models, future researchers can refine our simplistic scalar approach $z$ by which the individual transition probabilities are adjusted such that the resulting row sum in the transition matrix equals 1. We currently assume that $z$ is constant over time, though future work can certainly relax this assumption.
Alternatively, one may investigate the scaling approach from \citet{grimshaw2011markov}, who scaled the output from binary logistic regression models towards a similar goal.
Lastly, our study revealed that the Pearson residuals appear to be approximately normally distributed, though future work can focus on the theoretical grounding of these results; particularly since the distributional shape seems to be unknown.

Some further imitations are noted as follows. 
Our empirical analysis consumes a single dataset of South African residential mortgages, and the underlying portfolio is large and diverse. However, some of our results may differ when using data from other product types (e.g., credit cards), geographies, or banks with risk appetites and/or credit policies that are materially different. Replicating our study on other datasets may therefore be of interest.
Secondly, we assume that the observed historical relationships amongst macroeconomic, loan-level, and portfolio-level variables remain stable into the future in predicting transition probabilities. However, this assumption may be violated by structural economic shifts, regulatory changes, or novel borrower behaviours; such as those observed during COVID-19. Subsequently, future work can investigate the effect of regime switches in these relationships.
Thirdly, while our selected modelling techniques span useful points on the spectrum of complexity, it is a non-exhaustive list. Future work can explore other techniques, such as survival models with competing risks or machine learning approaches -- all of which may capture non-linearities even more effectively.
Despite these study limitations, we believe that our contributions advance the current practice in multistate PD-modelling, which can help produce more timeous and accurate ECL-estimates under IFRS 9.


\appendix
\section{Appendix}
\label{sec:Appendix}

In \autoref{app:betaRegression_basics}, we review the basic of beta regression (BR) models and its extension, BR with variable dispersion, which was ultimately used in this study. The application of Cook's distance is discussed in \autoref{app:cookDistance} towards removing influential observations from the samples on which BR-models are trained. Then, details of a goodness-of-fit analysis are given in \autoref{app:pearsonResiduals} using the Pearson residuals of BR-models. In \autoref{app:multiLogisReg_basics}, we discuss the fundamentals of multinomial regression logistic (MLR) models in predicting a multi-category and unordered outcome variable. A brief description of the set of input variables is given in \autoref{app:inputSpace}, as used across both the BR-- and MLR-models. Finally, we provide a list of acronyms in \autoref{app:acronyms}.

\subsection{The basics of beta regression}
\label{app:betaRegression_basics}

Introduced by \citet{ferrari2004beta}, a beta regression model relates a set of $p$ input variables $\boldsymbol{x}=\left\{ \boldsymbol{x}_1, \dots, \boldsymbol{x}_p \right\}$ to an outcome variable $y\in(0,1)$ that follows a beta distribution. 
These outcomes are typically asymmetrically distributed and heteroscedastic (i.e., observations are scattered closer to the mean than usual), which implies that inference based on homoscedasticity might be flawed. Conversely, a beta distribution is flexible and can contend with this asymmetry and heteroscedasticity by assuming a wide variety of distributional shapes.
One might be temped to transform the $y$-values into reals, followed by simply modelling the resulting mean thereof as a linear predictor $\eta$ of $\boldsymbol{x}$. However, the authors noted that it becomes awkward to interpret the subsequent model parameters in terms of the original outcome variable, at least relative to the ease thereof with beta regression. These reasons augur well for using beta regression in modelling transition rates as a function of $\boldsymbol{x}$, as in our context.


\citet{ferrari2004beta} provided a newly-formulated probability density function $f$ of the underlying random variable $Y\in(0,1)$ that is beta distributed, expressed as
\begin{equation} \label{eq:density_beta2}
    f(y,\mu, \phi) = \frac{\Gamma(\phi)}{\Gamma(\mu\phi) + \Gamma((1-\mu)\phi)}y^{\mu\phi-1}(1-y)^{(1-\mu)\phi - 1} \, , \quad \text{with} \ y \in(0,1) \, , \mu\in(0,1), \ \text{and} \ \phi>0 \, ,
\end{equation}
where $\Gamma(\cdot)$ is the Gamma function, and $\mu,\phi$ are shape parameters.
Accordingly, the mean and variance is respectively written as $\mathbb{E}(Y)=\mu$ and $\mathbb{V}(Y)=\mu(1-\mu)(1+\phi)^{-1}$.
The authors explain that greater values of the precision $\phi$ will correlate with a smaller variance in $Y$, having fixed the mean $\mu$ to some value.
Moreover, the density function $f$ from \autoref{eq:density_beta2} can assume a wide variety of shapes, even becoming symmetric for $\mu=0.5$, as shown in \citet{ferrari2004beta} across different values of $\mu$ and $\phi$.

Regarding the model form, let $Y_1,\dots,Y_n$ be a sample of independent random variables such that each $Y_i, i=1,\dots,n$ follows the same beta density from \autoref{eq:density_beta2}, albeit differently parametrised with mean $\mu_i$ and unknown precision $\phi$. The overall mean of $Y_i$ is then assumed as
\begin{equation} \label{eq:genericModelForm_beta}
    g_1(\mu_i) = \sum_{u=1}^{p_1}{\beta_{u}^\mathrm{T} x_{iu} = \eta_{1i}} \, , \nonumber
\end{equation}
where the vector $\boldsymbol{\beta}=\left\{\beta_{1}, \dots, \beta_{p_1} \right\}^\mathrm{T}$ contains $p_1$ estimable regression coefficients, $\eta_{1i}$ is the linear predictor, and $\boldsymbol{x}_i = \left\{x_{i1},\dots,x_{ip_1} \right\}$ are corresponding observations for subject $i$ from the inputs $\boldsymbol{x}$. The link function $g_1(\cdot)$ for the mean is strictly monotonic and twice-differentiable, and maps the response $(0,1)$ into a real value $\mathbb{R}$. A particularly popular choice for $g_1$ is the logit link $g(\mu)=\log{\left\{ \mu/(1-\mu)  \right\}}$, largely due to its relationship with the odds ratio in logistic regression, as well as its potential to improve the model fit. \citet{ferrari2004beta} noted that other choices include: 1) the probit link $g_1(\mu)= \Theta^{-1}(\mu)$ with $\Theta(\cdot)$ denoting the cumulative distribution function of a standard normal random variable; 2) the complementary log-log link $g(\mu)=\log{ \left\{ -\log{(1-\mu)} \right\}}$; and 3) the log-log link $g(\mu)=\log{ \left\{ -\log{(\mu)} \right\}}$.
In obtaining estimates for $\boldsymbol{\beta}$ and $\phi$, the authors derived the log-likelihood function from \autoref{eq:density_beta2} for a logit link function. 
Under regularity conditions, \citet{ferrari2004beta} explained that the expected value of the derivative of the log-likelihood function will be zero. Hence, one may write $\mathbb{E}(y_i^{*}) = \mu_i^*=\psi\left( \mu_i\phi \right) - \psi\left( (1-\mu_i)\phi \right)$ with $y_i^{*}=\log{\left(y_i/(1-y_i)\right)}$, where $\psi(\cdot)$ is the Digamma function; i.e., $\psi(z)=\mathrm{d}\log{(\Gamma(z))}/\mathrm{d}z$.
Finally, the log-likelihood is differentiated with respect to each unknown parameter, whereafter the resulting score functions (respective to $\boldsymbol{\beta}$ and $\phi$) are set to zero. Doing so enables a numerical procedure to maximise the log-likelihood function in practice, thereby obtaining estimates for $\boldsymbol{\beta}$ and $\phi$.

In extending the BR-model, \citet{simas2010improved} formulated the \textit{variable dispersion beta regression} (VDBR) model by restructuring the precision parameter $\phi$ within a regression framework, similar to the mean parameter $\mu$. In so doing, one can embed the outcome variable's heteroscedasticity via a series of input variables that are specific to the precision parameter. The precision (or "variance function") of $Y_i$ may then be similarly modelled as
\begin{equation} \label{eq:genericModelForm_beta_precision}
    g_2(\phi_i) = \sum_{u=1}^{p_2}{\theta_{u}^\mathrm{T} z_{iu} = \eta_{2i} } \, , \nonumber
\end{equation}
where the vector $\boldsymbol{\theta}=\left\{\theta_{1}, \dots, \theta_{p_2} \right\}^\mathrm{T}$ contains $p_2$ unknown regression coefficients,  $\eta_{2i}$ is the linear predictor of subject $i$, and $\boldsymbol{z}_i = \left\{z_{i1},\dots,z_{ip_2} \right\}$ are corresponding observations from the inputs $\boldsymbol{z}$, which may overlap with $\boldsymbol{x}$. Both link functions $g_1(\cdot): (0,1) \rightarrow \mathbb{R}$ and $g_2(\cdot): (0,\infty) \rightarrow \mathbb{R}$ are assumed to be strictly monotonic and twice-differentiable, as achieved respectively by the logit and the log functions, amongst others.
Thereafter, the authors showed that the parameters $\boldsymbol{\beta}$ and $\boldsymbol{\theta}$ are simultaneously obtained by maximising the associated log-likelihood function, similar to \citet{ferrari2004beta}. Once estimated, one can produce predictions as $\hat{y}_i=\hat{\mu}_i\approx \mathbb{E}(Y_i)$ while the estimated variance thereof is $s^2(y_i)=\hat{\mu}_i(1-\hat{\mu}_i)/(1+\hat{\phi}_i) \approx \mathbb{V}(Y_i)$, where $\hat{\mu}_i=g_1^{-1}(\boldsymbol{\beta}^\mathrm{T} \boldsymbol{x}_i)$ and $\hat{\phi}_i=g_2^{-1}( \boldsymbol{\theta}^\mathrm{T} \boldsymbol{z}_i )$. While $\hat{\phi}_i$ does not directly influence the prediction $\hat{y}_i$, it does affect the degree to which it can vary, particularly since $\boldsymbol{\beta}$ and $\boldsymbol{\theta}$ influence each other during the estimation procedure.

\subsection{Removing influential observations using Cooks' distance}
\label{app:cookDistance}

Given its popularity, we use Cook's distance to identify influential observations within the sample $\mathcal{D}_T$ from which BR-models are trained, thereby improving the model fit upon the removal of these observations. From \citet{ferrari2004beta}, Cook's distance $D_\mathrm{C}(i)$ measures the influence of the $i$th observation on the estimated regression coefficients $\hat{\boldsymbol{\beta}}$ against those coefficients $\hat{\boldsymbol{\beta}}_{(i)}$ that were estimated without $i$ in the sample. More formally, $D_\mathrm{C}(i)$ is the squared distance between $\hat{\boldsymbol{\beta}}$ and $\hat{\boldsymbol{\beta}}_{(i)}$, and its calculation would usually require fitting the BR-model $n+1$ times. Instead, \citet{ferrari2004beta} explained that $D_\mathrm{C}(i)$ may be approximated by calculating
\begin{equation} \label{eq:cookDistance}
    D_\mathrm{C}(i) = \frac{h_{ii}r_i^2}{p(1-h_{ii})^2}.
\end{equation}
In \autoref{eq:cookDistance}, $p$ is the number of regression coefficients in $\hat{\boldsymbol{\beta}}$, $r_i$ is the Pearson residual (see \autoref{app:pearsonResiduals}) between the observed response $y_i$ and the fitted value $\hat{y}_i$, and $h_{ii}$ is the leverage of $y_i$ from the hat matrix; itself derived specially for beta regression by \citet{ferrari2004beta}.

\begin{figure}[ht!]
    \centering\includegraphics[width=0.8\linewidth,height=0.45\textheight]{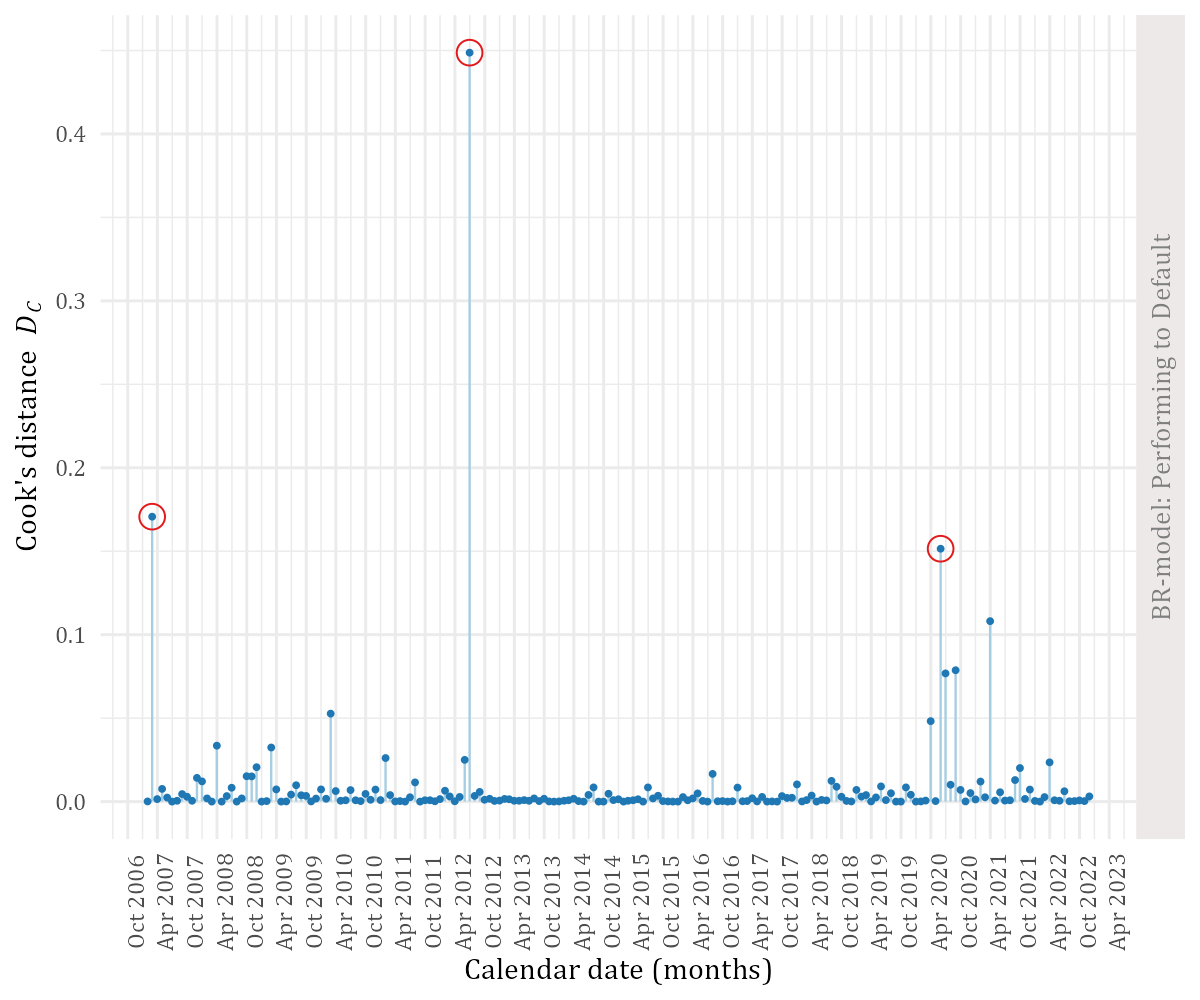}
    \caption{Cook's distance plot for the BR-model of transition type P$\rightarrow$D. Encircled points indicate highly influential observations.}\label{fig:Cook}
\end{figure}

Cook’s distance $D_\mathrm{C}$ is subsequently used to tweak the BR-models by identifying and removing a few influential observations, which generally improved the prediction accuracy of the models. In the interest of brevity, we shall only report and discuss the $D_\mathrm{C}$-plot for the P$\rightarrow$D transition type and its corresponding BR-model, though similar results hold for the other BR-models; see the codebase from \citet{botha2025sourcecode}. \autoref{fig:Cook} shows $D_\mathrm{C}(i)$ for each monthly observation $i$, and three observations are identified as relatively influential: the transition rates for Jan-2007, Jun-2012, and Jul-2012. We experimented with removing different combinations of these influential observations across all BR-models and, using the MAE as measure, obtained the best fit when generally removing the Jun-2012 observation. In \autoref{tab:CookMAE}, we summarise the change in the MAE-measure upon removing the influential observations from $\mathcal{D}_T$. In the vast majority of cases, the removal yielded a superior fit in the underlying BR-models, albeit small -- except for the P$\rightarrow$S transition type, in which case the opposite is true. Nonetheless, even a small improvement in the model fit can yield a significant financial impact on large loan portfolios. We therefore appreciate the removal of influential observations as a crucial step in building BR-models, especially when dealing with such small sample sizes.

\begin{table}[ht!]
    \centering
    \caption{Pseudo $R^2_\mathrm{F}$-values before and after deleting influential observations identified using Cook's distance $D_\mathrm{C}$ from \autoref{eq:cookDistance}, before applying any scaling of the output.}
    \label{tab:CookMAE}
    \begin{tabular}{lcccccc}
        \toprule
         \textbf{Transition type $kl$:}&  P$\rightarrow$P& P$\rightarrow$D & P$\rightarrow$S & D$\rightarrow$D & D$\rightarrow$S & D$\rightarrow$W \\
         \midrule
         \textbf{Before:}& 63.85\% & 86.46\% & 47.11\% & 32.06\% &  58.21\% & 39.23\%\\
         \textbf{After:}&  69.96\% & 87.21\% & 60.08\% & 33.27\% & 63.47\% & 39.74\%\\
        \bottomrule
    \end{tabular}
\end{table}

\subsection{Analysing the Pearson residuals of the BR-models in gauging their goodness-of-fit}
\label{app:pearsonResiduals}

It is standard practice in statistical modelling to analyse the residuals $r_i=y_i-\hat{y}_i$ between a model's predictions $\hat{y}_i$ and the outcomes $y_i$ for observations $i=1,\dots,n$, thereby assessing the model's goodness-of-fit to its training data $\mathcal{D}_T$. However, these `raw' residuals of BR-models are not typically used given that the outcomes (and the resulting BR-models) are generally heteroscedastic, as discussed by \citet{ferrari2004beta} and \citet{cribari2010beta}. Instead, these authors suggested that one should use \textit{Pearson} residuals (also called \textit{standardised ordinary} residuals), which are calculated as 
\begin{equation} \label{eq:PearsonRes}
    r_\mathrm{i}^{(\mathrm{P})} = \frac{y_i-\hat{y}_i}{\sqrt{\hat{s}^2(y_i)}} = \frac{y_i-\hat{y}_i}{\sqrt{ \hat{\mu}_i(1-\hat{\mu}_i)/(1+\hat{\phi}_i  }}\, , \nonumber
\end{equation}
where $\hat{s}^2(y_i)$ is the estimated variance across all $y_i$, as described in \autoref{app:betaRegression_basics}.
\citet{ferrari2004beta} admitted that the distribution of $r_i^{(\mathrm{P})},i=1,\dots,n$ is not exactly known, though we reasonably expect these residuals to follow a standard normal distribution. Accordingly, one can identify distinct patterns of outlying residuals or distributional shapes that are asymmetric, which ordinarily suggests a poor fit.

\begin{figure}[ht!]
    \centering
    \begin{subfigure}{0.49\textwidth}
        \caption{Performing to Performing} \label{fig:HistoPearson_PP}
        \centering\includegraphics[width=1\linewidth,height=0.27\textheight]{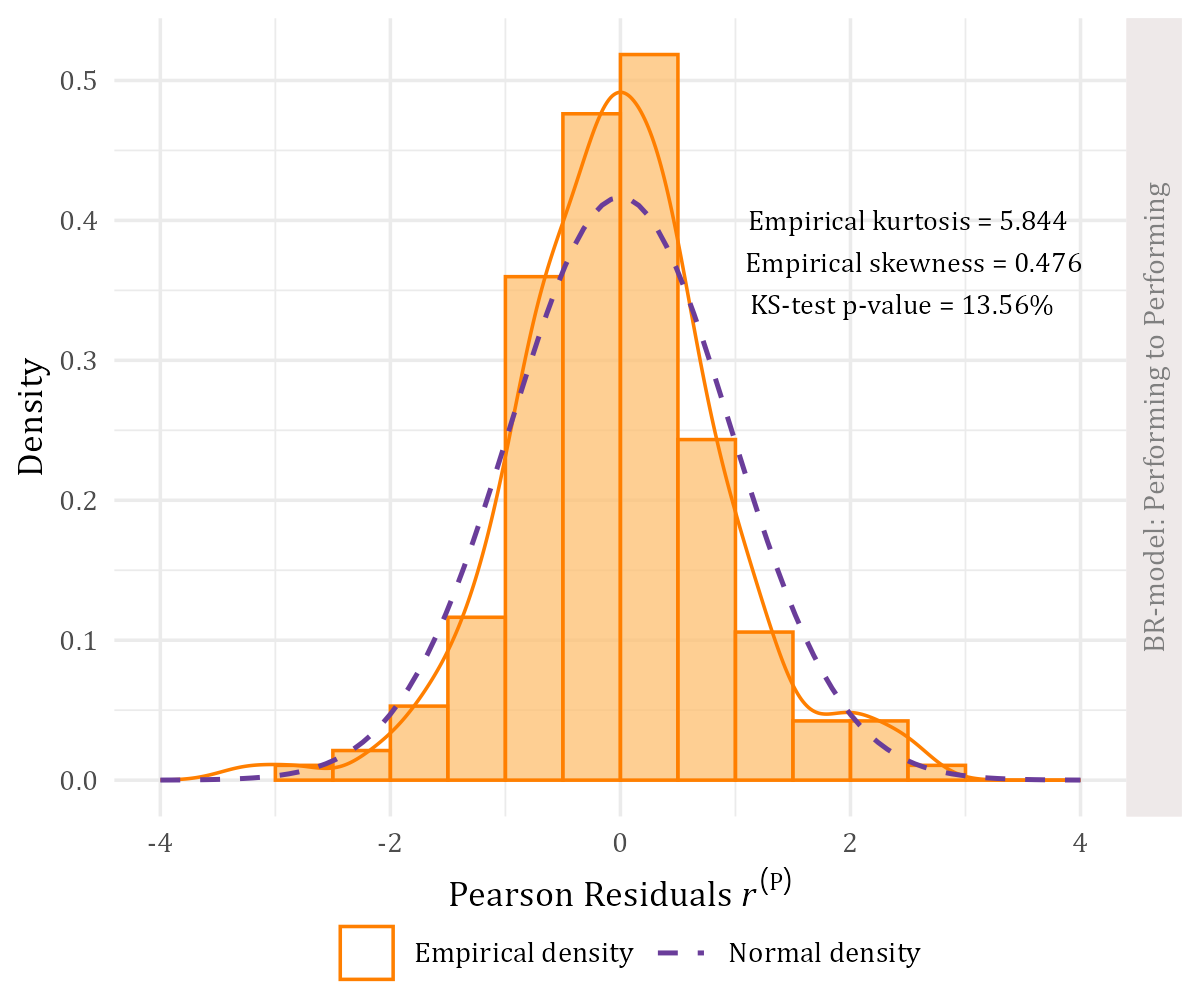}
    \end{subfigure}
    \begin{subfigure}{0.49\textwidth}
        \caption{Performing to Default} \label{fig:HistoPearson_PD}
        \centering\includegraphics[width=1\linewidth,height=0.27\textheight]{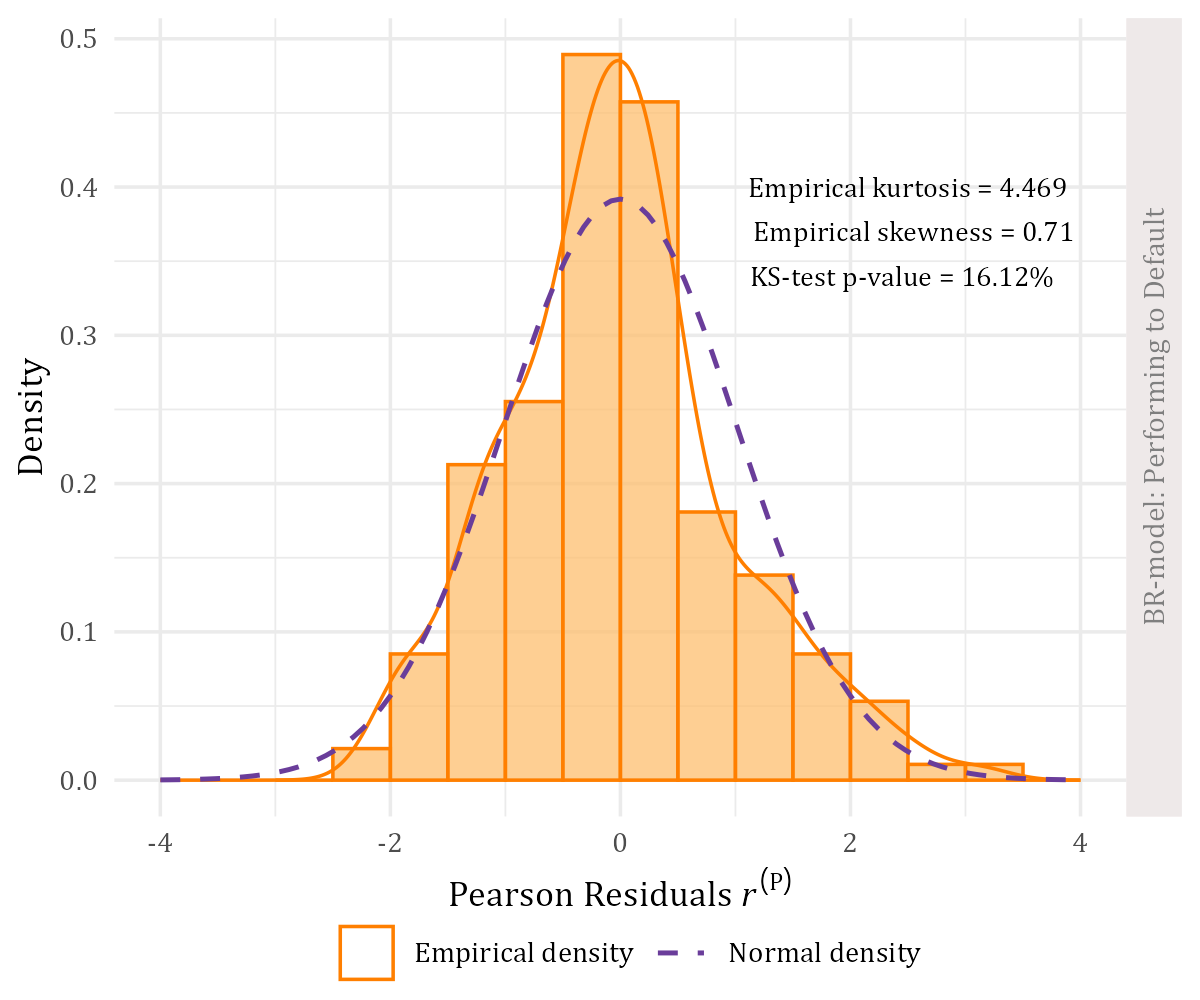}
    \end{subfigure} \\
    \begin{subfigure}{0.49\textwidth}
        \caption{Performing to Settlement} \label{fig:HistoPearson_PS}
        \centering\includegraphics[width=1\linewidth,height=0.27\textheight]{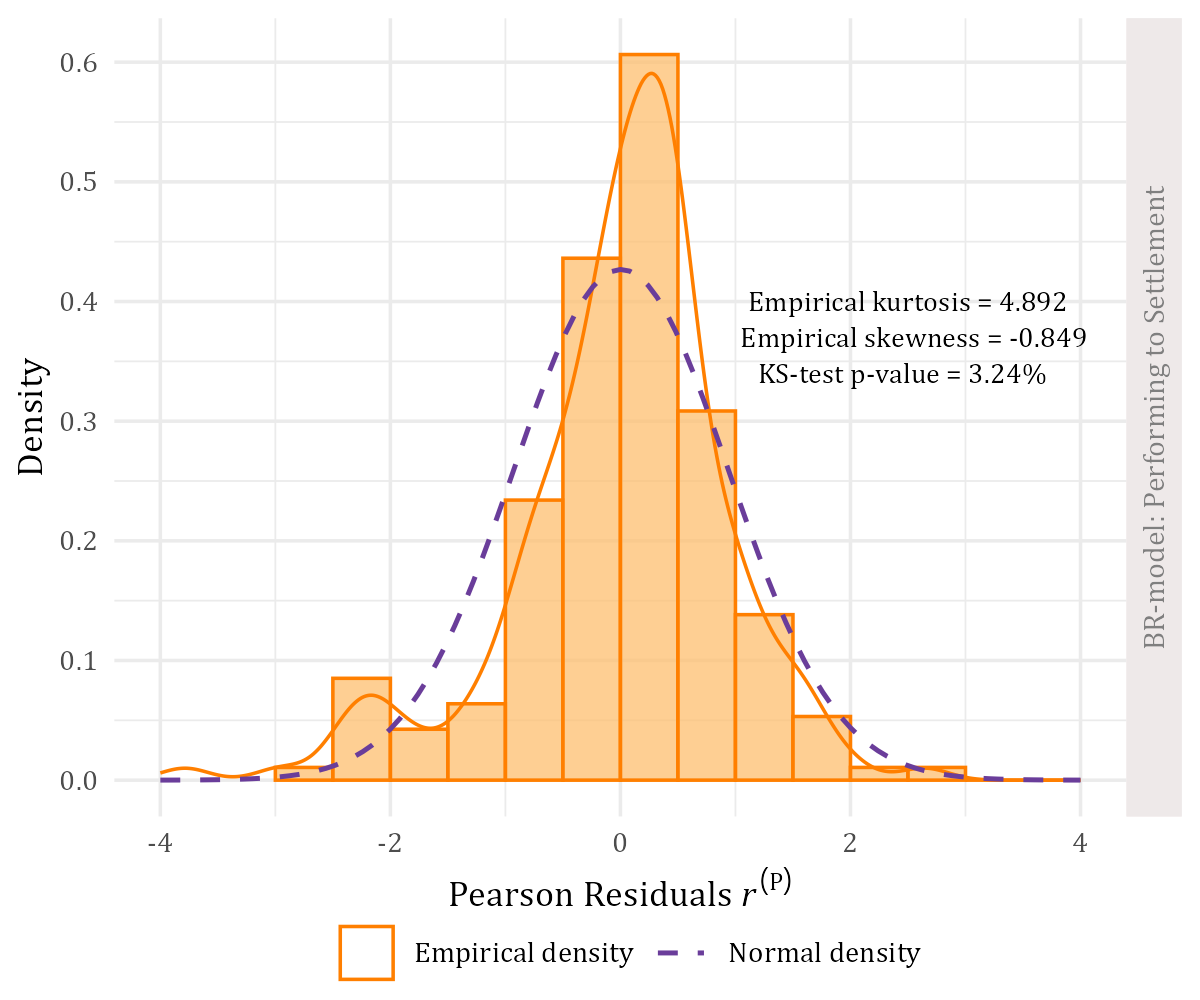}
    \end{subfigure}
    \begin{subfigure}{0.49\textwidth}
        \caption{Default to Default} \label{fig:HistoPearson_DD}
        \centering\includegraphics[width=1\linewidth,height=0.27\textheight]{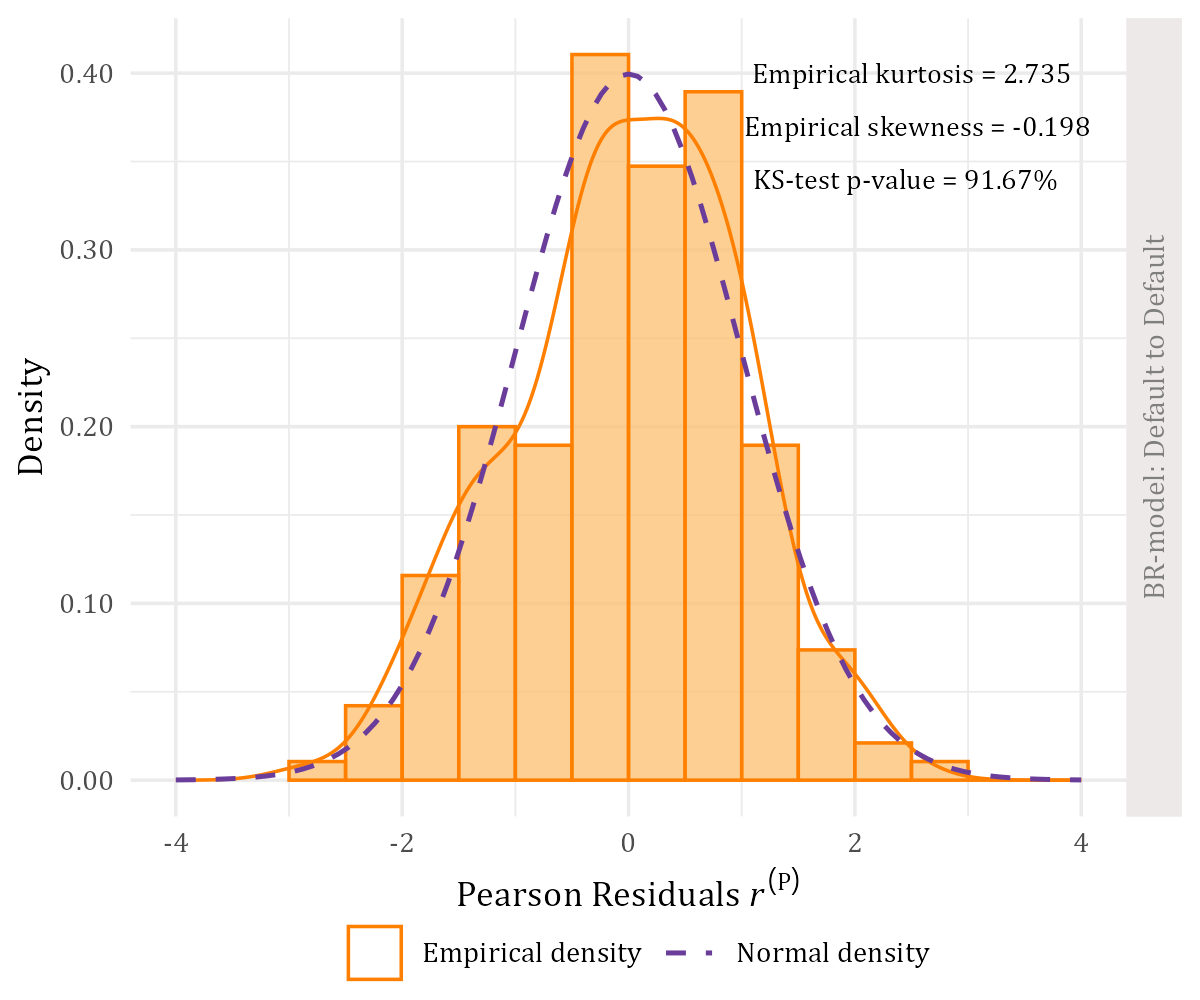}
    \end{subfigure} \\
    \begin{subfigure}{0.49\textwidth}
        \caption{Default to Settlement} \label{fig:HistoPearson_DS}
        \centering\includegraphics[width=1\linewidth,height=0.27\textheight]{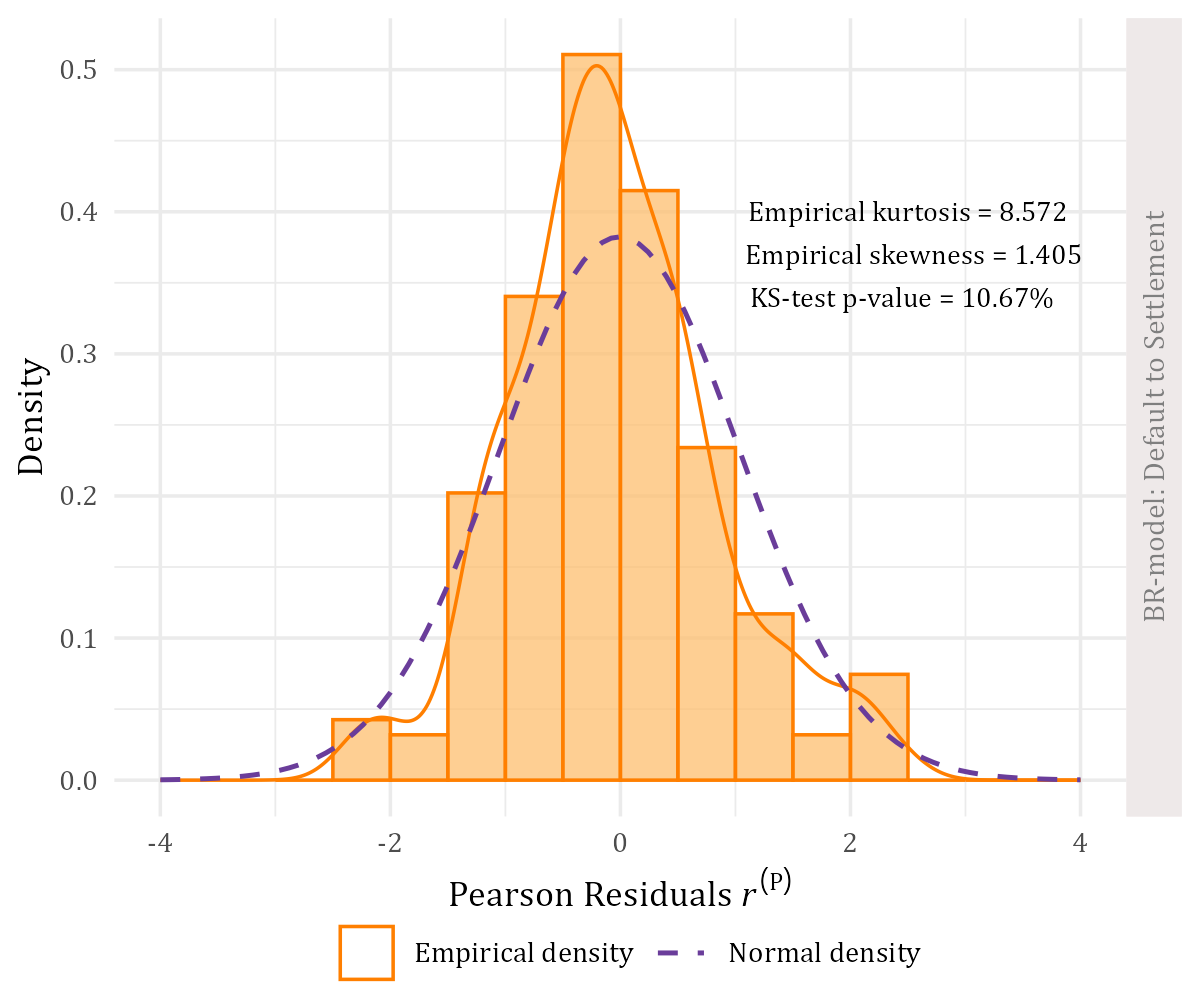}
    \end{subfigure}    
    \begin{subfigure}{0.49\textwidth}
        \caption{Default to Write-off} \label{fig:HistoPearson_DW}
        \centering\includegraphics[width=1\linewidth,height=0.27\textheight]{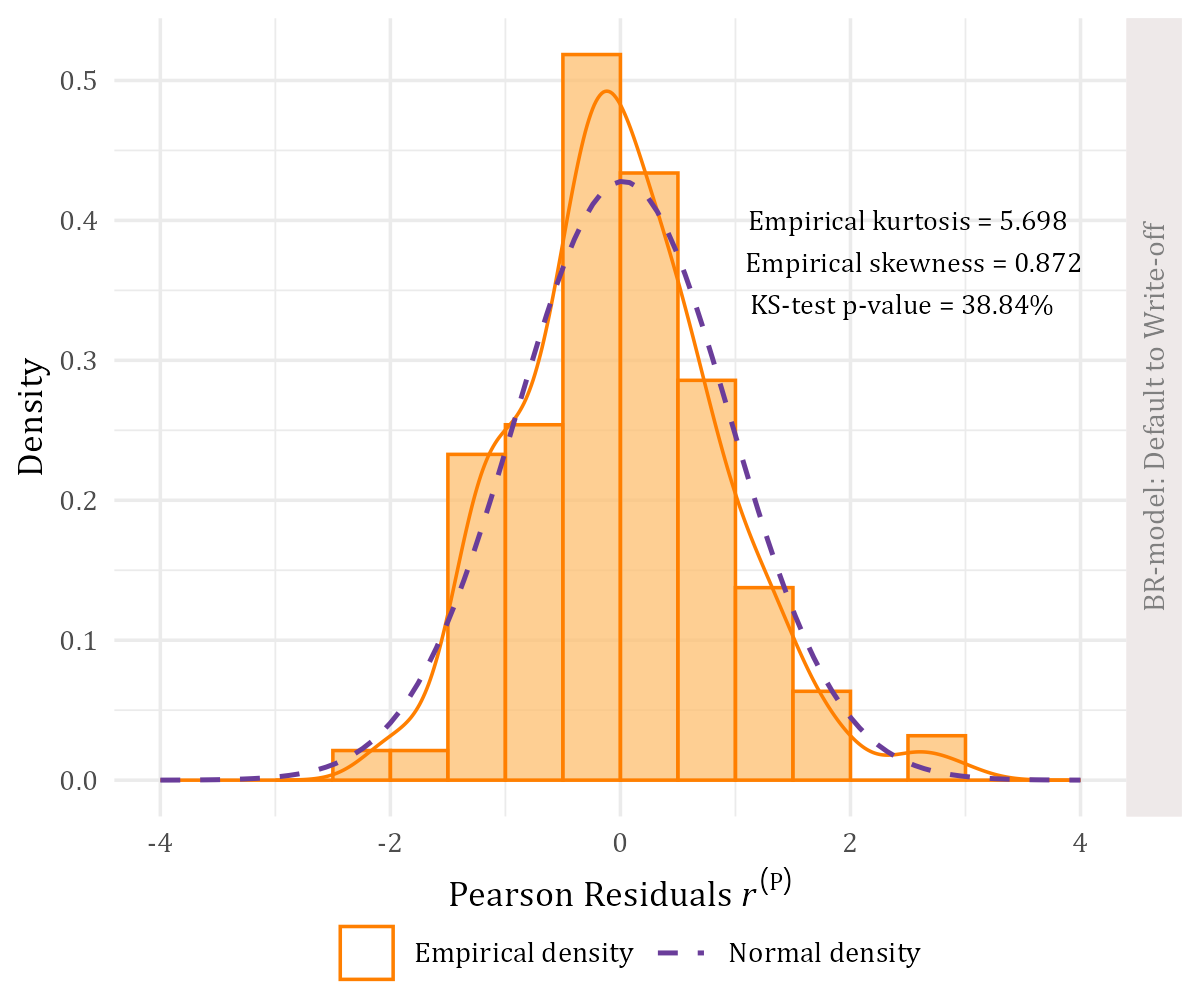}
    \end{subfigure}        
    \caption{Histograms and empirical densities of Pearson residuals $r_\mathrm{i}^{(\mathrm{P})}$ in gauging the fit of the BR-models.}\label{fig:HistogramPearson}
\end{figure}

For each of the BR-models, we compare the distribution of Pearson residuals to a standard normal distribution, as shown in  \autoref{fig:HistogramPearson}. With the exception of \autoref{fig:HistoPearson_PS}, the residual distributions appear to be mostly symmetric with close proximity to the normal distributions. Of the various transition types, the residual distribution of D$\rightarrow$S has the greatest skewness-value, based on the widely-used Fisher-Pearson skewness coefficient, as described by \citet{doane2011Skewness}. In addition to a graphical analysis, we also conduct a formal test of normality, having used the one-sample \textit{Kolmogorov-Smirnov} (KS) test. However, the KS-test failed to reject the null hypothesis (of normality) at the significance level of $\alpha=5\%$ for most distributions, which suggests that there is little statistical difference between the residual distributions and the standard normal distribution, despite visual analysis. Put differently, it would appear that the Pearson residuals follow approximately at least a symmetric distribution of some kind. We note however that other tests of normality might find differently, though such tests are outside of our scope.

\subsection{The fundamentals of multinomial logistic regression}
\label{app:multiLogisReg_basics}

As a generalisation of binary logistic regression, the response variable $Y$ within a \textit{multinomial logistic regression} (MLR) model is nominal, and can assume $J>2$ unordered possible outcomes. From \citet[\S 8.1]{hosmer2000logistic} and \citet[\S 1.2.2, \S 6.1]{agresti2007categorical}, the probabilities $\{ p_1, \dots, p_j, \dots, p_J \}$ of assuming any particular outcome $j$ can be written in terms of the category counts $n_1, \dots, n_j, \dots, n_J$. These category counts follow a multinomial distribution, which yields the joint probability of assuming any particular combination of category counts. An MLR-model then relates the conditional probability $p_j(\boldsymbol{x}_j)=\mathbb{P}\left(Y=j \, | \, \boldsymbol{x} \right)$ to a set of input variables $\boldsymbol{x}_j$ for category $j$ using a logit link function $g(\cdot)$. In particular, and with reference to some baseline-category $J'\in[1,J]$, the conditional mean $\mu_{ji}$ for the $j$th outcome and for loan $i=1,\dots,n$ is modelled as
\begin{equation} \label{eq:genericModelForm_MLR}
    g(\mu_{ji}) = \log{\left( \frac{p_j(\boldsymbol{x}_{ji})}{p_j(\boldsymbol{x}_{J'i})} \right)} = \eta_{ji} \quad \text{for } j\ne J' \, ,
\end{equation}
where $\eta_{ij}=\beta_{j0} + \beta_{j1} x_{ji1} + \beta_{jp} x_{jip}$ is the linear predictor of $p$ input variables, and $\boldsymbol{\beta}_j = \left\{ \beta_{j0}, \beta_{j1}, \dots, \beta_{jp} \right\}$ is a vector of estimable regression coefficients.

The formulation in \autoref{eq:genericModelForm_MLR} implies that an MLR-model will have $J-1$ logit functions, where each model $j$ has a separate coefficient vector $\boldsymbol{\beta}_j$ from the next model. In turn, one will have to estimate $(J-1)(p+1)$ coefficients in total.
The MLR coefficients $\boldsymbol{\beta}=\left\{ \boldsymbol{\beta}_1, \dots, \boldsymbol{\beta}_{J-1} \right\}$ are estimated simultaneously by means of maximising the conditional log-likelihood, which \citet[\S 8.1]{hosmer2000logistic} defined as
\begin{equation} \label{eq:conditionalLikelihood} 
    l(\boldsymbol{\beta})=\prod_{j=1}^{J} \prod_{i=1}^{n} p_{ji}(\boldsymbol{x}_{ji})^{{y}_{ji}} \, . \nonumber
\end{equation}
Finally, the probabilities $p_j(\boldsymbol{x}_{ij})$ from \autoref{eq:genericModelForm_MLR} given inputs $\boldsymbol{x}_{ij}$ can be expressed as 
\begin{equation} \label{eq:genericModelForm_MLR_probs} 
    p_{ji}\left(\boldsymbol{x}_{ji} \right) = \frac{ \exp{\left( \eta_{ji} \right)}}{ \sum_{u=1}^J{ \exp{\left( \eta_{ui} \right)} } } \quad \text{for } j\ne J'   \quad \text{and} \quad p_{ji}\left(\boldsymbol{x}_{ji} \right) = \frac{1}{\sum_{u=1}^J{ \exp{\left( \eta_{ui} \right)} }}\, \quad \text{for } j=J' \, . \nonumber
\end{equation}

An MLR-model imposes what is known as the \textit{Independence of Irrelevant Alternatives} (IIA) assumption. Following \citet{mcfadden1972conditional}, this assumption asserts that, for any two alternative categories $a_1\in \{1,\dots,J\}$ and $a_2\in \{1,\dots,J\}$ with $a_1 \neq a_2$, the relative odds ratio $\mathbb{P}(Y=a_1 \mid \boldsymbol{x})/\mathbb{P}(Y=a_2 \mid \boldsymbol{x})$ is invariant to the presence or characteristics of any other alternative $a_3\ne a_1,a_2$. Within the MLR-model, this assumption follows directly from the functional form
\begin{equation}
    \frac{\mathbb{P}(Y=a_1 \mid \boldsymbol{x})} {\mathbb{P}(Y=a_2 \mid \boldsymbol{x})} = \exp{\left( \boldsymbol{x}^\mathrm{T}(\boldsymbol{\beta}_{a_1}-\boldsymbol{\beta}_{a_2}) \right)} \, .
\end{equation}
As such, the relative odds depend only on the systematic components that drive each category (and not on latent factors), as captured by the coefficient vector $\boldsymbol{\beta}_j$ of each outcome $j$. Introducing or removing an additional category $a_3$ should not change how likely $a_1$ is chosen relative to $a_2$.
As a canonical illustration, consider predicting the type of vehicle that someone may choose when travelling: a red bus, a car, or a train. 
Suppose there is a latent factor, such as the passenger's comfort, that influences the eventual vehicle choice. Under the IIA-assumption, adding a blue bus option implies that the probability mass should primarily be drawn from the original bus category, rather than from the car or train categories. This is because the latent factor affecting bus choice is assumed to be independent of (or uncorrelated with) the other alternatives.

\subsection{A description of selected input variables}
\label{app:inputSpace}

In \autoref{tab:featuresDescription}, we describe the set of selected input variables for each BR-- and MLR-model, following our thematic variable selection process (as described in the main text). We provide only a high-level description of each variable, together with its particular selection(s) into specific models, whilst omitting the coefficient estimates in the interest of brevity. That said, a few summary statistics are given in \autoref{tab:SummaryStats} of what we deem to be the most important variables.
Quite a few single-factor models are built as a natural consequence of our thematic selection process, which revealed a few key insights regarding prediction power. Firstly, the variable \texttt{g0\_Delinq} proved to be a major source of prediction power in both MLR-models, which suggests that it (or delinquency-themed variables for that matter) should be retained as a baseline in all MLR-based PD-modelling. Secondly, the macroeconomic variables provided only a marginal lift in the fit statistics (AIC and the McFadden $R^2_\mathrm{McF}$) and the AUC, though they still remained statistically significant in predicting transition probabilities.

Regarding the MLR-models, some input variables are entered into the model using \textit{splines} towards improving the model fit. A spline can help model non-linear effects between an input $X$ and the outcome $Y$, and produces a smooth flexible curve. From \citet{perperoglou2019review}, we define a set of knots $\tau_1<\cdots < \tau_K$ by which the range $[a,b]$ of $X$ is partitioned. A spline $f(X)$ is then a smooth function with polynomial degree $d$ (usually cubic with $d=3$), and is defined as a composite of a series of basis functions $B_1(X), B_2(X), \dots$, expressed as
\begin{equation} \label{eq:splineRepresent}
    f(X) = \sum_{k=1}^{K+d+1}{\beta_kB_k(X)} \, , \nonumber
\end{equation}
where $\beta=\left\{\beta_1,\dots,\beta_{K+d+1} \right\}$ are estimable spline coefficients. Whilst a few types of spline basis functions exist, we opted for natural cubic splines given their popularity and their linear smoothness at the boundary knots; see \citet{perperoglou2019review} for a more in-depth review.
We used the \texttt{ns()} function from the \texttt{splines} R-library with expert judgement and experimented with various choices of knot numbers until all variables are statistically significant at $\alpha=5\%$. These natural splines are applied on the variables listed in \autoref{tab:MLR_splines} per MLR-model, together with the chosen number of knots.

\begin{longtable}{p{3.7cm} p{9.3cm} p{0.9cm} p{1.9cm}}
\caption{The selected input variables across the various transition type models (BR and MLR). Subscripts $[\mathrm{a}]$ denote loan account-level variables, $[\mathrm{p}]$ are portfolio-level inputs, and $[\mathrm{m}]$ represent macroeconomic covariates.}
\label{tab:featuresDescription} \\
\toprule
\textbf{Variable} & \textbf{Description} & \textbf{Model-type} & \textbf{Model states $kl$} \\ 
\midrule
\endfirsthead
\caption[]{(continued)} \\
\toprule
\textbf{Variable} & \textbf{Description} & \textbf{Model-type} & \textbf{Model states $kl$} \\ 
\midrule
\endhead
\midrule \multicolumn{4}{r}{\textit{Continued on next page}} \\
\endfoot
\bottomrule
\endlastfoot
\multirow[t]{2}{*}{\footnotesize{\texttt{AgeToTerm\_Avg}$_{[\mathrm{a}]}$} } & \footnotesize{Mean value of the ratio between a loan's age and its term.} & BR & PS; DS; DW \\
& & MLR & D$l$ \\ 
\footnotesize{\texttt{ArrearsDir\_3}$_{[\mathrm{a}]}$}  & \footnotesize{The trending direction of the arrears balance over 3 months, obtained qualitatively by comparing the current arrears-level to that of 3 months ago, binned as: 1) increasing; 2) milling; 3) decreasing (reference); and 4) missing.} & MLR & P$l$; D$l$ \\
\footnotesize{\texttt{ArrearsToBalance\_Pc}$_{[\mathrm{p}]}$} & \footnotesize{The sum of arrears divided by the sum of outstanding balances.} & BR & DD  \\
\footnotesize{\texttt{BalanceToPrincipal}$_{[\mathrm{a}]}$} & \footnotesize{Outstanding balance divided by the principal (loan amount) of the loan.} & MLR & P$l$; D$l$  \\
\multirow[t]{2}{*}{\footnotesize{\texttt{CreditLeverage}$_{[\mathrm{p}]}$} } & \footnotesize{The ratio between the sum of all outstanding balances and the sum of all principals, as a measure of portfolio maturity.} & BR & PP; PS; DW \\
& & MLR & P$l$; D$l$ \\ 
\footnotesize{\texttt{Curing\_Pc}$_{[\mathrm{a}]}$} & \footnotesize{Fraction of the portfolio that have newly cured from default.} & BR & DD; DW \\
\footnotesize{\texttt{DefaultStatus\_Avg}$_{[\mathrm{p}]}$} & \footnotesize{Fraction of the portfolio in default.} & MLR & P$l$ \\
\footnotesize{\texttt{DefaultStatus\_Avg\_1}$_{[\mathrm{p}]}$} & \footnotesize{1-month lagged version of \texttt{DefaultStatus\_Avg}.} & BR & PD; DW \\
\footnotesize{\texttt{DefaultStatus\_Avg\_2}$_{[\mathrm{p}]}$} & \footnotesize{2-month lagged version of \texttt{DefaultStatus\_Avg}.} & BR & PD \\
\footnotesize{\texttt{DefaultStatus\_Avg\_5}$_{[\mathrm{p}]}$} & \footnotesize{5-month lagged version of \texttt{DefaultStatus\_Avg}.} & MLR & D$l$ \\
\footnotesize{\texttt{DefaultStatus\_Avg\_6}$_{[\mathrm{p}]}$} & \footnotesize{6-month lagged version of \texttt{DefaultStatus\_Avg}.} & BR & DD; DW \\
\footnotesize{\texttt{DefaultStatus\_Avg\_12}$_{[\mathrm{p}]}$} & \footnotesize{12-month lagged version of \texttt{DefaultStatus\_Avg}.} & BR & DD \\
\footnotesize{\texttt{g0\_Delinq}$_{[\mathrm{a}]}$} & \footnotesize{Delinquency measure: number of payments in arrears; see $g_0$-measure in \citet{botha2021paper1}. Factorised version.} & MLR & P$l$; D$l$ \\
\multirow[t]{2}{*}{\footnotesize{\texttt{g0\_Delinq\_Avg}$_{[\mathrm{p}]}$} } & \footnotesize{Non-defaulted average delinquency $g_0$.} & BR & PP; DS \\
& & MLR & P$l$; D$l$ \\
\footnotesize{\texttt{g0\_Delinq\_Any\_Avg}$_{[\mathrm{p}]}$} & \footnotesize{Non-defaulted fraction of the portfolio with any degree of delinquency beyond $g_0=0$.} & BR & PP \\
\footnotesize{\texttt{g0\_Delinq\_Any\_Avg\_1}$_{[\mathrm{p}]}$} & \footnotesize{1-month lagged version of \texttt{g0\_Delinq\_Any\_Avg}.} & BR & PP \\
\footnotesize{\texttt{g0\_Delinq\_Any\_Avg\_2}$_{[\mathrm{p}]}$} & \footnotesize{2-month lagged version of \texttt{g0\_Delinq\_Any\_Avg}.} & BR & PS \\
\footnotesize{\texttt{g0\_Delinq\_1\_Avg}$_{[\mathrm{p}]}$} & \footnotesize{Fraction of the portfolio with $g_0=1$ payments in arrears.} & BR & PP \\
\multirow[t]{2}{*}{ \footnotesize{\texttt{g0\_Delinq\_2\_Avg}$_{[\mathrm{p}]}$}} & \footnotesize{Fraction of the portfolio with $g_0=2$ payments in arrears.} & BR & PD \\
 & & MLR & P$l$ \\
\footnotesize{\texttt{g0\_Delinq\_3\_Avg}$_{[\mathrm{p}]}$} & \footnotesize{Fraction of the portfolio with $g_0=3$ payments in arrears.} & BR & PS; DS \\
\footnotesize{\texttt{g0\_Delinq\_Num}$_{[\mathrm{a}]}$} & \footnotesize{Number of times that the $g_0$-measure has changed in value over loan life so far.} & MLR & P$l$; D$l$ \\
\footnotesize{\texttt{g0\_Delinq\_SD\_6}$_{[\mathrm{a}]}$} & \footnotesize{The sample standard deviation of \texttt{g0\_Delinq} over a rolling 6-month window.} & MLR & P$l$ \\
\footnotesize{\texttt{g0\_Delinq\_SD\_9}$_{[\mathrm{a}]}$} & \footnotesize{The sample standard deviation of \texttt{g0\_Delinq} over a rolling 9-month window.} & MLR & D$l$ \\
\footnotesize{\texttt{InterestRate\_Margin}$_{[\mathrm{a}]}$} & \footnotesize{Margin between a loan's nominal interest rate and the current prime lending rate, set by the South African Reserve Bank (SARB). Can be negative.} & MLR & P$l$; D$l$ \\
\footnotesize{\texttt{IntRate\_Margin\_Avg}$_{[\mathrm{p}]}$} & \footnotesize{The portfolio-level average of \texttt{InterestRate\_Margin} at each time point.} & BR & PD  \\
\multirow[t]{2}{*}{ \footnotesize{\texttt{M\_DebtToIncome}$_{[\mathrm{m}]}$} } &\footnotesize{Debt-to-Income: Average household debt expressed as a percentage of household income per quarter, interpolated monthly.}  & BR & PD \\
 & & MLR & D$l$ \\
\footnotesize{\texttt{M\_DebtToIncome\_1}$_{[\mathrm{m}]}$} &\footnotesize{1-month lagged version of \texttt{M\_DebtToIncome}.}  & BR & PD \\
\footnotesize{\texttt{M\_DebtToIncome\_12}$_{[\mathrm{m}]}$} &\footnotesize{12-month lagged version of \texttt{M\_DebtToIncome}.}  & BR & PW \\
\multirow[t]{2}{*}{ \footnotesize{\texttt{M\_Employment\_Growth}}$_{[\mathrm{m}]}$ } & \footnotesize{Year-on-year growth rate in the 4-quarter moving average of employment per quarter, interpolated monthly.} & BR & PS; DD; DW \\
 & & MLR & P$l$ \\
\footnotesize{\texttt{M\_Employment\_Growth\_1}}$_{[\mathrm{m}]}$  & \footnotesize{1-month lagged version of \texttt{M\_Employment\_Growth}.} & BR & PS \\
\footnotesize{\texttt{M\_Employment\_Growth\_9}}$_{[\mathrm{m}]}$  & \footnotesize{9-month lagged version of \texttt{M\_Employment\_Growth}.} & BR & PD \\
\footnotesize{\texttt{M\_Employment\_Growth\_12}}$_{[\mathrm{m}]}$  & \footnotesize{12-month lagged version of \texttt{M\_Employment\_Growth}.} & BR & PD; DW \\
\multirow[t]{2}{*}{ \footnotesize{\texttt{M\_Inflation\_Growth\_2}$_{[\mathrm{m}]}$}} & \footnotesize{Year-on-year growth rate in inflation index (CPI) per month, lagged by 2 months.} & BR & PD \\
& &  MLR & P$l$ \\
\footnotesize{\texttt{M\_Inflation\_Growth\_6}$_{[\mathrm{m}]}$} & \footnotesize{Year-on-year growth rate in inflation index (CPI) per month, lagged by 6 months.} & MLR & D$l$ \\ 
\footnotesize{\texttt{M\_RealGDP\_Growth}$_{[\mathrm{m}]}$} & \footnotesize{Year-on-year growth rate in the 4-quarter moving average of real GDP per quarter, interpolated monthly.} & MLR & D$l$ \\
\footnotesize{\texttt{M\_RealGDP\_Growth\_3}$_{[\mathrm{m}]}$}  & \footnotesize{3-month lagged version of \texttt{M\_RealGDP\_Growth}.} & BR & DW \\ 
\footnotesize{\texttt{M\_RealGDP\_Growth\_9}$_{[\mathrm{m}]}$}  & \footnotesize{9-month lagged version of \texttt{M\_RealGDP\_Growth}.} & BR & PP \\ 
\footnotesize{\texttt{M\_RealGDP\_Growth\_12}$_{[\mathrm{m}]}$}  & \footnotesize{12-month lagged version of \texttt{M\_RealGDP\_Growth}.} & BR & PP \\ 
\footnotesize{\texttt{M\_RealIncome\_Growth}$_{[\mathrm{m}]}$}  & \footnotesize{Year-on-year growth rate in the 4-quarter moving average of real income per quarter, interpolated monthly.} & BR & DS \\
\footnotesize{\texttt{M\_RealIncome\_Growth\_1}$_{[\mathrm{m}]}$}  & \footnotesize{1-month lagged version of \texttt{M\_RealIncome\_Growth}.} & BR & DS \\
\footnotesize{\texttt{M\_RealIncome\_Growth\_9}$_{[\mathrm{m}]}$}  & \footnotesize{9-month lagged version of \texttt{M\_RealIncome\_Growth}.} & BR & PP \\
\footnotesize{\texttt{M\_RealIncome\_Growth\_12}$_{[\mathrm{m}]}$}  & \footnotesize{12-month lagged version of \texttt{M\_RealIncome\_Growth}.} & BR & PP \\
\multirow[t]{2}{*}{ \footnotesize{\texttt{M\_Repo\_Rate}$_{[\mathrm{m}]}$} } & \footnotesize{Prevailing repurchase (or policy) rate set by the South African Reserve Bank (SARB).} & BR & PP; PD; PS\\
 & & MLR & P$l$; D$l$ \\
 \footnotesize{\texttt{M\_Repo\_Rate\_12}$_{[\mathrm{m}]}$}  & \footnotesize{12-month lagged version of \texttt{M\_Repo\_Rate}.} & BR & DD \\
\footnotesize{\texttt{NewLoans\_Pc\_3}$_{[\mathrm{p}]}$} & \footnotesize{Fraction of the portfolio that constitutes new loans, lagged by 3 months.} & BR& PP \\
\footnotesize{\texttt{PayMethod}$_{[\mathrm{a}]}$} & \footnotesize{A categorical variable designating different payment methods: 1) debit order (reference); 2) salary; 3) payroll or cash; and 4) missing.} & MLR & P$l$; D$l$ \\
\footnotesize{\texttt{PerfSpell\_Maturity\_Avg}$_{[\mathrm{p}]}$}  & \footnotesize{Mean value of performance spell ages at a particular point in (calendar) time.} & BR & PP; PS; DS; DW \\
\footnotesize{\texttt{Principal\_Real}$_{[\mathrm{a}]}$} & \footnotesize{Inflation-adjusted principal loan amount.} & MLR & D$l$ \\
\footnotesize{\texttt{Prepaid\_Pc}$_{[\mathrm{a}]}$} & \footnotesize{The prepaid or undrawn fraction of the available credit limit.} & MLR & P$l$ \\
\footnotesize{\texttt{Prev\_DS}$_{[\mathrm{p}]}$}  & \footnotesize{Previous transition rate for the transition type D$\rightarrow$S.} & BR & DS \\
\footnotesize{\texttt{Prev\_DW}$_{[\mathrm{p}]}$}  & \footnotesize{Previous transition rate for the transition type D$\rightarrow$W.} & BR & DW \\
\footnotesize{\texttt{RollEver\_24}$_{[\mathrm{a}]}$} & \footnotesize{Number of times that loan delinquency increased during the last 24 months, excluding the current time point.} & MLR & P$l$; D$l$ \\
\footnotesize{\texttt{StateSpell\_Num\_Total}$_{[\mathrm{a}]}$} & \footnotesize{The current state spell number, or total number of visits across all states over loan life.} & MLR & P$l$ \\
\footnotesize{\texttt{TimeInDelinqState}$_{[\mathrm{a}]}$} & \footnotesize{Duration (in months) of current delinquency `state' (or value) before the $g_0$-measure changes again in \texttt{g0\_Delinq} to another value.} & MLR & D$l$ \\ 
\footnotesize{\texttt{TimeInStateSpell}$_{[\mathrm{a}]}$} & \footnotesize{Duration (in months) spent so far in the current state.} & MLR & D$l$ \\ 
\end{longtable}

\begin{table}[!ht]
    \centering
    \caption{Selected variables on which natural regression splines are fit, along with the number of knots.} \label{tab:MLR_splines}
    \begin{tabular}{llc}
        \toprule
         \textbf{Variable} & \textbf{MLR model} $kl$ & \textbf{Knots} \\
         \midrule
         \texttt{BalanceToPrincipal} & P$l$ & 3 \\
         \texttt{CreditLeverage} & P$l$ & 3 \\
         \texttt{g0\_Delinq\_Num} & P$l$ & 5 \\
         \multirow[t]{2}{*}{\texttt{InterestRate\_Margin} } & P$l$ & 3 \\
         & D$l$ & 3 \\
         \texttt{M\_Repo\_rate} & P$l$ & 3 \\
         \multirow[t]{2}{*}{\texttt{RollEver\_24}} & P$l$ & 5 \\
         & D$l$ & 6 \\         
         \texttt{Prepaid\_Pc} & P$l$ & 4 \\
         \bottomrule
    \end{tabular}
\end{table}

\begin{table}[!ht]
    \centering
    \caption{Summary statistics of selected variables within the BR-- and MLR-models: mean, standard deviation (sd), and selected quantiles (10\%, 25\%, 50\%, 75\%, 90\%).} \label{tab:SummaryStats}
    \begin{tabular}{llllllll}
        \toprule
         \textbf{Variable} & Mean & sd & 10\% & 25\% & 50\% & 75\% & 90\% \\
         \midrule
         \texttt{AgeToTerm\_Avg} & 0.359 & 0.055 & 0.255 & 0.326 & 0.390 & 0.400 & 0.404 \\
         \texttt{BalanceToPrincipal} & 0.707 & 0.328 & 0.072 & 0.528 & 0.854 & 0.968 & 0.998 \\
         \texttt{CreditLeverage} & 0.766 & 0.031 & 0.725 & 0.745 & 0.756 & 0.795 & 0.815 \\
         \texttt{DefaultStatus\_Avg} & 0.051 & 0.015 & 0.039 & 0.041 & 0.048 & 0.057 & 0.077 \\
         \texttt{g0\_Delinq} & 0.188 & 0.635 & 0.000 & 0.000 & 0.000 & 0.000 & 1.000 \\
         \texttt{g0\_Delinq\_Avg} & 0.061 & 0.025 & 0.043 & 0.045 & 0.050 & 0.061 & 0.107 \\
         \texttt{InterestRate\_Margin} & -0.007 & 0.012 & -0.019 & -0.015 & -0.008 & 0.000 & 0.006 \\
         \texttt{M\_Employment\_Growth} & 0.003 & 0.021 & -0.029 & -0.001 & 0.008 & 0.016 & 0.026 \\
         \texttt{M\_Repo\_Rate} & 0.065 & 0.021 & 0.040 & 0.055 & 0.062 & 0.070 & 0.100 \\
         \texttt{RollEver\_24} & 0.481 & 1.051 & 0.000 & 0.000 & 0.000 & 0.484 & 2.000 \\
         \bottomrule
    \end{tabular}
\end{table}
\newpage
\subsection{A list of acronyms used in this paper}
\label{app:acronyms}

\begin{description}[font=\normalfont\bfseries, leftmargin=2cm, labelwidth=1.5cm]
  \item[AIC] Akaike Information Criterion
  \item[BR] Beta Regression
  \item[D] Default state
  \item[ECL] Expected Credit Loss
  \item[GoF] Goodness-of-Fit
  \item[IFRS] International Financial Reporting Standards
  \item[IIA] Independence of Irrelevant Alternatives
  \item[KS] Kolmogorov-Smirnov
  \item[LGD] Loss Given Default
  \item[MAE] Mean Absolute Error
  \item[MRM] Model Risk Management
  \item[MLR] Multinomial Logistic Regression
  \item[OLS] Ordinary Least Squares
  \item[P] Performing state
  \item[PD] Probability of Default
  \item[SD] Standard deviation
  \item[S] Settlement state 
  \item[TTC] Through-The-Cycle
  \item[VDBR] Variable Dispersion Beta Regression
  \item[W] Write-off state
\end{description}



\singlespacing
\printbibliography 
\onehalfspacing



\end{document}